\documentclass[reprint,amsmath,amssymb,aps,superscriptaddress,nofootinbib,twocolumn,10pt]{revtex4-1}

\usepackage[colorlinks=true,allcolors=blue]{hyperref}
\usepackage{graphicx,color}
\usepackage{dcolumn}
\usepackage{subfigure}
\usepackage{bm}
\usepackage{amsmath,amsfonts,amssymb}
\usepackage{acronym}
\usepackage{enumitem}
\usepackage{array}
\usepackage{multirow}

\makeatletter
\def\l@subsubsection#1#2{}
\def\l@subsubsubsection#1#2{}
\makeatother

\allowdisplaybreaks
\newcommand{\be}{\begin{equation}}
\newcommand{\ee}{\end{equation}}
\newcommand{\bea}{\begin{eqnarray}}
\newcommand{\eea}{\end{eqnarray}}

\newcommand{\SOUTHCUT}{
School of Physics and Optoelectronics, South China University of Technology, Guangzhou 510641,
People's Republic of China}

\newcommand{\SYSU}{
School of Physics and Astronomy, Sun Yat-sen University (Zhuhai Campus), Zhuhai 519082, China}

\newcommand{\Sichuan}{
School of Physics and Electronic Engineering, Sichuan University of Science $\&$ Engineering, Zigong 643000, China}

\newacro{EMRI}{extreme mass-ratio inspiral}
\newacro{MBH}{massive black hole}
\newacro{BH}{black hole}
\newacro{GR}{general relativity}
\newacro{HKBH}{hairy Kerr black hole}
\newacro{KNBH}{Kerr-Newmann black hole}
\newacro{KBH}{Kerr black hole}
\newacro{NHT}{no-hair theorem}
\newacro{DWD}{double white dwarf}
\newacro{GW}{gravitational wave}
\newacro{AK}{analytic kludge}
\newacro{NK}{numerical kludge}
\newacro{AAK}{augmented analytic kludge}
\newacro{CO}{compact object}
\newacro{PE}{parameter estimation}
\newacro{SNR}{signal-to-noise ratio}
\newacro{PN}{post-Newtonian}
\newacro{FIM}{fisher information matrix}
\newacro{LSO}{last stable orbit}
\newacro{ISCO}{innermost stable circular orbit}
\newacro{BBH}{Binary Black Hole}
\newacro{BNS}{Binary Neutron Star}
\newacro{NS}{Neutron Star}
\newacro{KN}{Kerr-Newmann}
\newacro{ECO}{exotic compact object}
\newacro{MCO}{massive compact object}
\newacro{ECOs}{exotic compact objects}
\begin{document}
\title{Detecting the tidal heating with the generic extreme mass-ratio inspirals}

\author{Tieguang Zi}
\affiliation{\SOUTHCUT}

\author{Chang-Qing Ye}
\email{yechq6@mail2.sysu.edu.cn}
\affiliation{\Sichuan}
\affiliation{\SYSU}

\author{Peng-Cheng Li}
\email{corresponding author: pchli2021@scut.edu.cn}
\affiliation{\SOUTHCUT}

\begin{abstract}
The horizon of a classical black hole (BH), functioning as a one-way membrane, plays a vital role in the dynamic evolution of binary BHs, capable of absorbing fluxes entirely.
Tidal heating, stemming from this phenomenon, exerts a notable influence on the production of gravitational waves (GWs). If at least one member of a binary is an exotic compact object (ECO) instead of a BH, the absorption of fluxes is expected to be incomplete and the tidal heating would be different. Thus, tidal heating can be utilized for model-independent investigations into the nature of compact object.
In this paper, assuming that the \ac{EMRI} contains a stellar-mass compact object orbiting around a massive ECO  with a reflective surface, we compute the GWs from the generic EMRI orbits. Using the accurate and analytic flux formulas in the black hole spacetime, we adapted these formulas in the vicinity of the ECO surface by incorporating a reflectivity parameter.
Under the  adiabatic approximation, we can evolve the orbital parameters and compute the EMRI waveforms. The effect of tidal heating for the spinning and non-spinning objects can be used to constrain the reflectivity of the surface at the level of $\mathcal{O}(10^{-6})$ by computing the mismatch and fisher information matrix.
\end{abstract}
\maketitle

\section{Introduction}
In the framework of classical general relativity (GR), the event horizon of a \ac{BH}, as a null hypersurface, can absorb all radiation near the horizon.
The famous Hawking radiation stated that there would exist the non-trivial quantum effects due to the causal structure of the horizon \cite{Hawking:1976ra}.
From the viewpoint of quantum gravity, the spacetime structure near the horizon could be corrected at the Planck scale \cite{Bianchi:2020miz,Giddings:2017jts,Maselli:2017cmm,Agullo:2020hxe,Bianchi:2020miz,Bianchi:2020bxa}. There is another possibility that the extremely compact objects can be formed via the phase transition \cite{Mazur:2004fk,Mazur:2004fk}. All these configuration would result in the similar objects, where the external spacetimes of compact objects behave almost the same as BHs, except for that near the horizons. These BH-like compact objects are usually called as \ac{ECOs} \cite{Cardoso:2019rvt}.

So far, the nearly one hundred GW events emitted from the coalesces of compact binaries have been reported, which provided the novel ways to test GR and the chances to find the evidences beyond GR \cite{LIGOScientific:2018dkp,LIGOScientific:2020tif,LIGOScientific:2021djp,LIGOScientific:2021sio}. Nowadays ones can confirm that there indeed exist massive BHs
in the universe \cite{EventHorizonTelescope:2019dse,EventHorizonTelescope:2019dse,EventHorizonTelescope:2022wkp,EventHorizonTelescope:2019ggy}, their spacetime and dynamics can be well described by the Kerr metric according to the no-hair theorem \cite{Kerr:1963ud,Bekenstein:1996pn,Chrusciel:2012jk}.
 The Event Horizon Telescope (EHT) also provides a new means to probe the near-horizon physical effect \cite{EventHorizonTelescope:2019dse,Gralla:2019xty,EventHorizonTelescope:2022xqj}.
According to the null-hypothesis testings and searches for signatures of GR violations, no reliable evidence has been found in the observation data sets detected by LIGO, Virgo, and EHT \cite{LIGOScientific:2018dkp,LIGOScientific:2019fpa,LIGOScientific:2020tif,LIGOScientific:2021sio,
Ghosh:2022xhn,EventHorizonTelescope:2020qrl,EventHorizonTelescope:2022xqj}. It is very likely because  the current observed GW events have the relatively small signal-to-noise ratios (SNRs). The future space-based and ground-based gravitational wave detectors expected in the next decade, such as the Einstein Telescope \cite{Punturo:2010zz}, LISA \cite{LISA:2017pwj}, and TianQin \cite{TianQin:2015yph, Gong:2021gvw}, will be capable of observing various types of signals with higher SNRs originating from more massive gravitational sources. This capability will enable accurate tests of GR and the strong constraint of its alternatives in the strong field regime or near the horizon.

In Newtonian gravity, for a two-body system, orbital and rotational energy is dissipated as heat through the process of tidal friction, known as tidal heating. Similarly, for a classical black hole, the event horizon acts as a one-way membrane that can completely absorb incoming fluxes. This absorption by the BH's horizon can also be referred to as tidal heating \cite{Hartle:1973zz,Hughes:2001jr,Comeau:2009bz,Poisson:2014gka,Landry:2015zfa,Chatziioannou:2016kem,Pitre:2023xsr}. In addition, in Newtonian gravity, tidal heating for the rotating bodies is always accompanied by the transfer of energy and angular momentum from the bodies to their orbit.
For the rotating BHs case, this phenomenon would also happen.
Due to the mechanism of superradiance \cite{Brito:2015oca}, a rotating BH can not only absorb radiations but also amplify radiations of lower frequency. This then makes BHs become the dissipative systems, like the Newtonian viscous fluid \cite{Poisson:2009di,Cardoso:2012zn}.
During the inspiral of a binary BH, the radiation absorbed by one BH can be understood as the tidal field generated by the companion. This tidal interaction induces deformation on the horizon of the BH, especially at the late stage of the inspiral, which subsequently influences the companion's orbit, just like the bulge on the earth acts on motion of the moon \cite{Hughes:2001jr,OSullivan:2015lni}. As a result, as long as the BH rotates fast enough such that the superradiance condition is satisfied, energy and angular momentum are transferred from the BH into the orbit, contributing to the overall dynamics of the system \cite{Hartle:1973zz,Datta:2019epe}.
Assuming at least one member of a binary is an ECO, the absorption of radiation by this ECO could be significant reduced, depending on the reflectivity of its surface. Then the tidal heating and the inspirals in the binary would be also affected, which for the rapidly rotating objects might become more evident \cite{Datta:2019epe}. Therefore, tidal heating provides a model-independent test to discern the presence or absence of a horizon. With the incorporation of tidal heating, one expects that the exploration of physical effects at the horizon scale could be feasible through observations made by LISA \cite{Maselli:2017cmm,Johnson-Mcdaniel:2018cdu,Cardoso:2019rvt}.

An \ac{EMRI} consisting of a stellar-mass compact object (with mass $m\sim1-10^{2}M_\odot$)
 inspiraling around a \ac{MCO} (with mass $M\sim10^{5}-10^{9}M_\odot$) can emit the $m\rm Hz$ GW signals in the band of LISA, which can sever as an ideal tool to map the spacetime geometry of the central MCO \cite{Ryan:1995wh,Berry:2019wgg}.
To determine the nature of the MCO in the EMRI system, especially whether the horizon exists, various approaches have been proposed. If the MCO is an ECO instead of a BH, then the nonvanishing of tidal deformability would leave an  observable imprint on the EMRI waveform \cite{Pani:2019cyc,DeLuca:2022xlz,Zi:2023pvl}.  However, to our knowledge these works are limited to the nonrelativistic case, which may not be accurate to capture  the physics in the strong-field regime.  Moreover, tidal heating provides a model-independent tool to discern the presence of horizons for the MCOs in the EMRIs  \cite{Datta:2019euh,Datta:2019epe}.
In particular, Datta $et ~al.$ \cite{Datta:2019epe} argued that the tidal heating for the circular and equatorial EMRI orbits allows to constrain the reflectivity of the ECO at the level of $\mathcal{O}(0.01\%)$ by computing the mismatch between the relativistic waveforms from the Kerr BH and the ECO spacetimes \cite{Datta:2019epe}.
Further works were focused on modifying the boundary conditions near the surface of the ECOs. For example, Maggio  $et ~al.$ \cite{Maggio:2021uge} evaluated the bound on reflectivity using the EMRI around Kerr-like horizonless ECOs at the level of $\mathcal{O}(10^{-6})$, where the fluxes near the surface need to be modified
by imposing the ingoing and outgoing boundary conditions simultaneously.
Using the similar configuration, Sago $et ~al.$ found that there indeed exists the anomalous oscillating behaviors in the fluxes for the horizonless ECOs \cite{Sago:2021iku}, which is expected to be a signal of the presence of a reflective surface \cite{Sago:2022bbj}.

Note that the above analyses using the tidal heating within the EMRI dynamics to constrain the reflectivity of ECOs only focus on the equatorial and circular orbits. However, the typical EMRI systems are the inclined  and eccentric binary objects when they are birthed in the center of galaxies \cite{Berry:2019wgg}.
Therefore it is necessary and important to extend the work in  Ref.  \cite{Datta:2019epe} by
considering the generic EMRI orbits. Furthermore, it is also important to present the more rigorous constraint on the tidal heating parameter with the \ac{FIM}  \cite{Cutler:1994ys,Vallisneri:2007ev} rather than the mismatch.

We will assume that the MCO of the EMRI has a reflective surface instead of a horizon and the external spacetime is approximately described by the Kerr metric or Schwarzschild metric.
We will compute the parameterized fluxes near the surface in term of the analytic accurate flux formulas
for the Kerr BH \cite{Sago:2015rpa} and for the Schwarzschild BH \cite{Munna:2020juq,Munna:2023vds}.
Then the modified fluxes for the MCO can be embedded into a latest EMRI waveform model that can be used for data analysis, referred to as \texttt{FastEMRIWaveform} (\texttt{\texttt{FEW}}) \cite{Katz:2021yft} to generate the EMRI waveform rapidly.  In order to evaluate the effect of tidal heating on the EMRI orbital phase and waveform,
we will calculate dephasing and mismatch between waveforms from the ECO and Kerr or Schwarzschild BH.
Finally, to assess the capability of LISA for detecting the tidal heating effect,
we will place the constraint on the reflectivity using the \ac{FIM} method.

The arrangement of paper is formulated as follows. In Sec. \ref{Method}, we introduce the computation recipe of EMRI waveforms around the ECO  and GW data analysis, including
the spacetime of ECOs and the timelike geodesics in Sec. \ref{background},
the fluxes  in Sec. \ref{fluxes},
the evolution scheme of orbital parameters under the adiabatic
approximation in Sec. \ref{evolution} and  the method of EMRI waveform data analysis in Sec. \ref{waveformanalysis}.
In Sec. \ref{result}, we show the results of waveform comparison and the constraints on the reflectivity of the surface of the ECO using the EMRI waveforms produced by the \texttt{FEW} model. Sec. \ref{result:spinning} presents the findings for the spinning ECO case, while Sec. \ref{result:non-spinning} focuses on the non-spinning ECO case.
Finally, we present the conclusion in section \ref{Conclusion}. Throughout this paper, we use the geometric units $G=c=1$.
\section{SETUP}\label{Method}
\subsection{Background of the massive compact object and geodesics}\label{background}
In this subsection, we briefly show the external spacetime  of a rotating ECO \cite{Maggio:2017ivp,Maggio:2018ivz,Wang:2018gin}.
According to Birkhoff's theorem, the external spacetime of a non-spinning object is the case of
the Schwarzchild BH. When considering the full spacetime geometry of a rotating ECO, the Birkhoff's theorem would  be
invalid. However, in the higher compactness spacetime regime, the multipole moments
of a ECO can return to the case of a Kerr BH, this is due to the fact that the perturbative solutions of the vacuum
Einstein equation should obey the smooth BH limit \cite{Yagi:2015upa,Pani:2015tga,Raposo:2018xkf}.
Thus, we assume that the external spacetime of the ECO can be determined by the Kerr \ac{BH} and
the presence of a reflective surface, rather then the classical horizon.
The  Kerr metric in Boyer-Lindquist coordinates is given by
\begin{eqnarray}\label{KNQ:Boyer-Lindquist}
ds^2&=&-{\Delta\over\Sigma}(dt-a\sin^2\theta\,d\varphi)^2
+{\sin^2\theta\over\Sigma}\left[(r^2+a^2)d\varphi-a\,dt\right]^2
\nonumber\\
&&+{\Sigma\over\Delta}dr^2+\Sigma d\theta^2\,,
\end{eqnarray}
here $\Sigma=$ $r^2+a^2\cos^2\theta$ and $\Delta =$ $r^2-2Mr+a^2$.
The parameters $(M, a)$ represent for the mass and  spin of the ECO.

Firstly we introduce the geodesics of a timelike particle near the ECO. The orbits can be described by the Mino time $d\lambda$,
which is defined via proper time $d\tau$ by $d\lambda=d\tau/\Sigma$.
Then the geodesic equations of timelike particles are given by \cite{Thorne:1973zz,Drasco:2005kz},
\begin{eqnarray}
\left( \frac{dr}{d\lambda} \right)^2 &= &R, \label{eq:drdlam} \\
\left(\frac{d\theta}{d\lambda}\right)^2 &=& \Theta, \label{eq:dthetadlam}\\
\frac{dt}{d\lambda} &=& T_\theta + T_r,\\
\frac{d\phi}{d\lambda} &=& P_\theta + P_r,
\end{eqnarray}
with
\begin{eqnarray}
R &=& [E(r^2+a^2)-aL_z]^2 - \Delta[r^2+Q+(L_z-a E)^2], \\
\Theta &=& Q - \left[a^2(1-E^2) + \frac{L^2_z}{\sin^2\theta} \right]\cos^2\theta, \\
T_r &=& \frac{(r^2+a^2)^2}{\Delta} E - \frac{2MraL_z}{\Delta}, \\
T_\theta &=& -a^2\sin^2\theta E, \\
P_r &=& \frac{2MarE}{\Delta} -\frac{a^2L_z}{\Delta},\\
P_\theta &=& \csc^2\theta L_z,
\end{eqnarray}
where $E$, $L_z$ is the energy and angular momentum the orbit per unit mass $m$, respectively, and  $Q$ is the Carter constant of the orbit per unit mass square $m^2$.
Their analytic expressions and the fundamental orbital frequencies have been derived in Refs. \cite{Schmidt:2002qk,Fujita:2009bp}.

From the Eqs. \eqref{eq:drdlam} and \eqref{eq:dthetadlam},
the orbits oscillate in the ranges  $r_{\rm min}\leq r \leq r_{\rm max}$ and $\theta_{\rm min}\leq \theta \leq \theta_{\rm max}$ with $\theta_{\rm max}  = \pi - \theta_{\rm min}$ and $r_{\rm min/max}=p/(1\pm e)$, where
$p$ is the orbital semi-latus rectum and $e$ is the orbital eccentricity,
and the orbital inclination angle $I$ can be obtained by $\theta_{\rm min}$
\cite{Drasco:2005kz,Hughes:2021exa}, defined by $\theta_I=\pi/2 - \mathrm{sgn}  (L_z) \theta_{\rm min}$.
The parameter $x_I=\cos \theta_I$ changes from $-1$ to $1$ when the orbits transform from the retrograde equatorial to the prograde equatorial.

\subsection{Fluxes of ECOs}\label{fluxes}
In our hypothesis, the surface has an indirect influence on the orbits around the ECO and only modifies the fluxes near the surface.
The EMRI trajectories are a series of geodesics of the Kerr BH, whose evolution are subjected with fluxes formulas obtained by solving Teukolsky equation \cite{Hughes:2021exa,Isoyama:2021jjd}.
However, it takes a high computational cost to obtain adiabatic EMRI waveforms when computing the
FIM, this is mainly due to the longtime duration and the higher harmonic element of the EMRI waveforms for the generic orbits, the detailed discussion can refer to the paper \cite{Hughes:2021exa}.
The analytic \ac{PN} formulas of energy and angular fluxes have been derived
up to the 4PN order in the Kerr BH \cite{Sago:2015rpa} and to the higher order in the Schwarzschild BH \cite{Munna:2020juq, Munna:2023vds}.
If both the orbital velocity and  eccentricity are small, the relative errors of the analytic PN formulas can reach to the tolerable level \cite{Sago:2015rpa,Munna:2020juq, Munna:2023vds,Isoyama:2021jjd}.

Following the method of Ref. \cite{Datta:2019epe}, the fluxes absorbed by the surface of the ECO can be parameterized as following
\begin{equation}\label{hor:flux:eco}
\mathcal{\dot{C}}^H_{\rm ECO} = (1-|\mathcal{R}|^2)\mathcal{\dot{C}}^H_{\rm Kerr}
\end{equation}
where $\mathcal{C}\in \{E,L_z,Q\}$  and $\dot{\mathcal{C}}$ denote the energy flux, angular momentum flux and Carter constant fluxes near the surface of the ECO. The parameter $\mathcal{R}$ denotes the reflectivity coefficient of surface \cite{Maggio:2017ivp,Mark:2017dnq}.
For simplicity, we assume that $\mathcal{R}$ is a constant quantity independent of  the frequency. The parameter  $\mathcal{R}$ vanishes for a BH, however, it can be equal to one for a fully reflective surface of a ECO. On the other hand, for the fluxes at infinity, we make the hypothesis that the fluxes at infinity for the ECO  keep the same as those of  the Kerr BH.

Given that the geodesic constants $E$, $L_z$ and $Q$ are changing due to the presence of gravitational radiation, we can assume that the smaller object evolves from one geodesic to another under the adiabatic approximation.
The change of the orbital constants  are subjected to the fluxes radiated by the EMRI system.
The total fluxes of the EMRI orbits, according to the balance law \cite{Hughes:2021exa}, consist of the fluxes at the infinity and near the surface
\begin{eqnarray}\label{changerates}
\frac{d\mathcal{C}}{dt}^{\rm orbit} = -\left(\frac{d\mathcal{C}^{ \infty} }{dt}
+\frac{d\mathcal{C}^{H} }{dt}\right).
\end{eqnarray}
Here the  $\left(\frac{d\mathcal{C}}{dt}\right)^{\rm \infty}$ and $\left(\frac{d\mathcal{C}}{dt}\right)^{ H}$ denotes to the fluxes at infinity  and near the surface of the ECO. Note that the energy flux at infinity depends on the homogeneous solutions of Teukolsky equation that is regular at the horizon, which is different for an ECO. However, as shown in \cite{Datta:2019epe} the energy flux at infinity is, up to numerical accuracy,
the same for a BH or for an ECO, regardless of the reflectivity of the latter.

\subsection{Orbital adiabatic evolution}\label{evolution}
In this subsection, we introduce the evolution scheme of orbital parameters in the adiabatic approximation condition. The geodesics near the ECO can be parameterized by the orbital parameters $p$, $e$ and $x_I$,
which can also be transformed as expressions of the geodesic constants $E$, $L_z$ and $Q$ \cite{Hughes:2021exa}, then the two sets of quantities can both describe the geodesics equivalently.
 Therefore, we can evolve the orbital parameters $p$, $e$ and $x_I$ adiabatically using the
change rates of geodesic constants.

The translation formulas between the the orbital parameters and geodesic constants have been derived in Ref. \cite{Hughes:2021exa}
\begin{eqnarray}
\frac{d x_I }{dt}& =& \frac{2x_I(1-x_I^2)L_z \dot{L_z} - x^3_I\left[\dot{Q} + 2(1-x_I^2)a^2 E \dot{E}\right]}
{2\left[L_z^2 +x^4_I a^2(1-E^2) \right]}, \label{eq:xdot}\\
\frac{d e}{dt} &=&\frac{(1-e^2)}{2p} \left[(1-e)\frac{dr_a}{dt} - (1+e)\frac{dr_p}{dt}\right] ,\label{eq:pdot} \\
\frac{dp}{dt} &=&\frac{(1-e)^2}{2} \frac{dr_a}{dt} + \frac{(1+e)^2}{2}\frac{dr_p}{dt} ,\label{eq:edot}
\end{eqnarray}
where the expression $\frac{dr_{a,p}}{dt}$ is given by
\begin{equation}
\frac{dr_{a,p}}{dt} = J_{E_{r_{a,p}}} \dot{E} + J_{L_z{_{r_{a,p}}}} \dot{L_z} +  J_{Q_{r_{a,p}}} \frac{dQ}{dt}
\end{equation}
with
\begin{eqnarray}
J_{E_{r_{a,p}}}  & =& \frac{4aM(L_z -a E)r_{a,p} - 2E r_{a,p}^2 (a^2+r_{a,p}^2)}{\mathcal{D}(r_{a,p})}, \label{eq:JE}\\
J_{L_z{_{r_{a,p}}}} & =& \frac{4aM(L_z -a E)r_{a,p} + 2L_z r_{a,p}^2 }{\mathcal{D}(r_{a,p})}, \label{eq:JE}\\
J_{Q_{r_{a,p}}} &=& \frac{r_{a,p}^2 +2M r_{a,p} +a^2 }{\mathcal{D}(r_{a,p})}, \\
\mathcal{D}(r_{a,p}) &=& 2M\left[Q + (L_z-aE)^2\right]- 4r^3(1-E^2)
\\ &-&  2r \left[L_z^2 + Q + (1-E^2)a^2\right]  +6Mr^2.
\end{eqnarray}
Here the expressions $\{E,L_z,Q\}$ and $\{\dot{E},\dot{L_z},\dot{Q}\}$ are the geodesic constants and their variation rates.
The analytic expressions of geodesic constants $\{E,L_z,Q\}$ have been derived in Ref. \cite{Schmidt:2002qk,vandeMeent:2019cam} and their change rates (\ref{changerates}).

Using the analytic flux formulas for the Kerr \cite{Sago:2015rpa} and  Schwarzschild BHs
\cite{Munna:2020juq,Munna:2023vds}, we can obtain the trajectories of general EMRI orbits by
solving the ordinary differential equations \eqref{eq:xdot}-\eqref{eq:edot}, where the cutoff of orbital parameters evolution is taken as the \ac{LSO} of Kerr spacetime \cite{Stein:2019buj}.
\subsection{Waveform computation and analysis}\label{waveformanalysis}
The \texttt{FEW} model is a fast waveform generation tools for the future EMRI data analysis, which includes the eccentric adiabatic orbits for the Schwarzschild BH and the
generic adiabatic orbits  at 5PN order for the Kerr BH \cite{Katz:2021yft,michael_l_katz_2023_8190418,Chua:2017ujo,Speri:2023jte}.
We compute the waveforms in the \texttt{FEW} model using the inspiral trajectories of EMRI modified by the tidal heating effect as introduced in the subsection \ref{evolution}. With the EMRI trajectories produced by the \texttt{FEW} model, we can assess the effect of tidal heating on waveform phase by computing the dephasing of EMIR waveforms.  The waveform dephasing can be defined by the integral  of the differences of orbital frequencies between the  Kerr BH case and the ECO case
\bea
\delta \Phi_{i} = \int^T_0 (\Omega_{i}^{\mathcal{R}\neq 0} - \Omega_{i}^{\mathcal{R}=0})dt,
\eea
where $\Omega_{i} \in \{\Omega_{r},\Omega_{\theta}, \Omega_{\phi}\}$ are the orbital fundamental frequencies
that are derived by Refs. \cite{Schmidt:2002qk,Fujita:2009bp}. The superscripts $\mathcal{R}=0$
and $\mathcal{R}\neq0$ correspond to the frequencies computed with the trajectories of the Kerr BH and the ECO, respectively. A useful criterion for the distinguishing additional effects on waveforms has been proposed in Refs.\cite{Datta:2019epe, Speeney:2022ryg}, where the effect from tidal heating can be discerned if the dephasing satisfies $\delta \Phi_{i}\geq 1$. Note that as pointed out by Ref. \cite{Datta:2019epe} the dephasing only provides a very rough evaluation for the excluding parameters of tidal heating and a more careful analysis is needed.

To better assess the effect of tidal heating on EMRI waveforms  with space-borne GW detectors, we perform the mismatch analysis using two kinds of waveform where one is the Kerr waveform and the other is the ECO's.  For two waveforms $h_a(t)$ and $h_b(t)$, the mismatch $\mathcal{M}$ is defined by
\be\label{mis:defines}
 \mathcal{M}=1- \frac{<h_a|h_b>}{\sqrt{<h_a|h_a><h_b|h_b>}},
\ee
and the noise-weighted inner product $<*|*>$ is defined by
\begin{align}\label{inner}
<h_a |h_b > =2\int^\infty_0 df \frac{\tilde{h}_a^*(f)\tilde{h}_b(f)+\tilde{h}_a(f)\tilde{h}_b^*(f)}{S_n(f)},
\end{align}
where the quantities with tildes denote the Fourier transform of waveform data,
the star signs mean complex conjugation,  and $S_n(f)$ is noise power spectral density of LISA~\cite{LISA:2017pwj}.
The \ac{SNR} can be defined by
\begin{align}\label{inner}
\rho = \sqrt{<h |h >}.
\end{align}

From the definition of mismatch in Eq. \eqref{mis:defines}, one can see  that the mismatch is zero for two identical waveforms.
In this paper, we adopt a useful rule of thumb to distinguish between two waveforms in LISA observations. LISA can distinguish the differences in waveforms if the mismatch satisfies $\mathcal{M} \geq 1/(2\rho^2)$ \cite{Flanagan:1997kp,Lindblom:2008cm}.
The \ac{SNR} threshold  for EMRI signals detected by the LISA or TianQin is chosen as $\rho=20$ \cite{Babak:2017tow,Fan:2020zhy},
so the threshold value of mismatch is $\mathcal{M}_{\rm min}=0.001$. Note that when calculating the mismatch we only vary the reflectivity and let all other parameters fixed. Since the those other parameters will be unknown a priori, this may lead to a higher mismatch than the realistic case.

To assess the capability of future space-based GW detectors, such as LISA, for measuring the parameter $|\mathcal{R}|^2$ of tidal heating, we perform the parameter estimation using statistic methods in this paper.
The \ac{FIM} is a common method to estimate the measure error of GW source parameters, which can be
defined by
\bea
\Gamma_{ij} = \left(\frac{\partial h}{\partial \theta_i } | \frac{\partial h}{\partial \theta_j }\right)
\eea
where  $\theta_i$, $i=1,2,3,\cdots$, are the parameters of the modified \texttt{few} model describing EMRI systems with ECO.
The intrinsic parameters are classified as
$\theta_{\rm intr}  = (M,m,a,p_0,e_0,x_I, |\mathcal{R}|^2, \Phi_\phi^0, \Phi_\theta^0, \Phi_r^0)$,
where the quanties $(\Phi_\phi^0, \Phi_\theta^0, \Phi_r^0)$ represent for the initial values of azimuthal, polar and radial phases,
and the extrinsic parameters are $\theta_{\rm extr} = (D_L,\vartheta_S,\varphi_S, \vartheta_K,\varphi_K)$.
$D_L$ refers to the luminosity distance from the source to the detector,
and $(\vartheta_S,\varphi_S, \vartheta_K, \varphi_K)$ are the  azimuth and colatitude of source sky position and orbital angular momentum.
For the EMRI signal with higher SNR, the variance-covariance matrix $\mathbf{\Sigma}$ can be given by the inverse of FIM
\bea
<\delta \theta^i \delta\theta^j > \backsimeq (\mathbf{\Gamma}^{-1})^{ij}=\mathbf{\Sigma}^{ij},
\eea
the uncertainty of $i-\rm th$ parameter $\theta_i$ can be
obtained by the square root of the variance-covariance matrix
\bea
\Delta \theta_i = \sqrt{<(\delta \theta_i )^2>} \simeq \sqrt{(\mathbf{\Sigma}_{ii})}.
\eea
It should be noted that the precondition of FIM method is linear signal approximation,
whose applicability has been argued in Ref. \cite{Zi:2022hcc}.
In principle, the FIM method is more rigorous and reliable than the previous two method to assess the impact of tidal heating on the waveform. For example, the correlation in parameters can be described by the off-diagonal elements of the inverse FIM. Moreover, the FIM can embody the impact of the reflectivity on the measurement accuracy of the other parameters.


Additionally, following Refs. \cite{Babak:2017tow,Fan:2020zhy}, we also compute the uncertainties of  solid angles by combining the measuring errors of the angles $(\vartheta_S,\varphi_S, \vartheta_K,\varphi_K)$ as
\bea
\Delta \Omega_i = 2\pi |\sin \vartheta_i| \sqrt{\mathbf{\Sigma}^2_{\vartheta_i}\mathbf{\Sigma}^2_{\varphi_i} - \mathbf{\Sigma}^2_{\vartheta_i \varphi_i} }
\eea
with $i\in\{S,K\}$.

\section{Result}\label{result}
In this section we compare the EMRI waveforms from the ECO and the Kerr BH, and compute the constraint on the reflectivity $|\mathcal{R}|^2$ using the FIM method.

\subsection{Spinning massive objects}\label{result:spinning}
\begin{figure*}[th]
 \centering
 \includegraphics[width=0.95\textwidth]{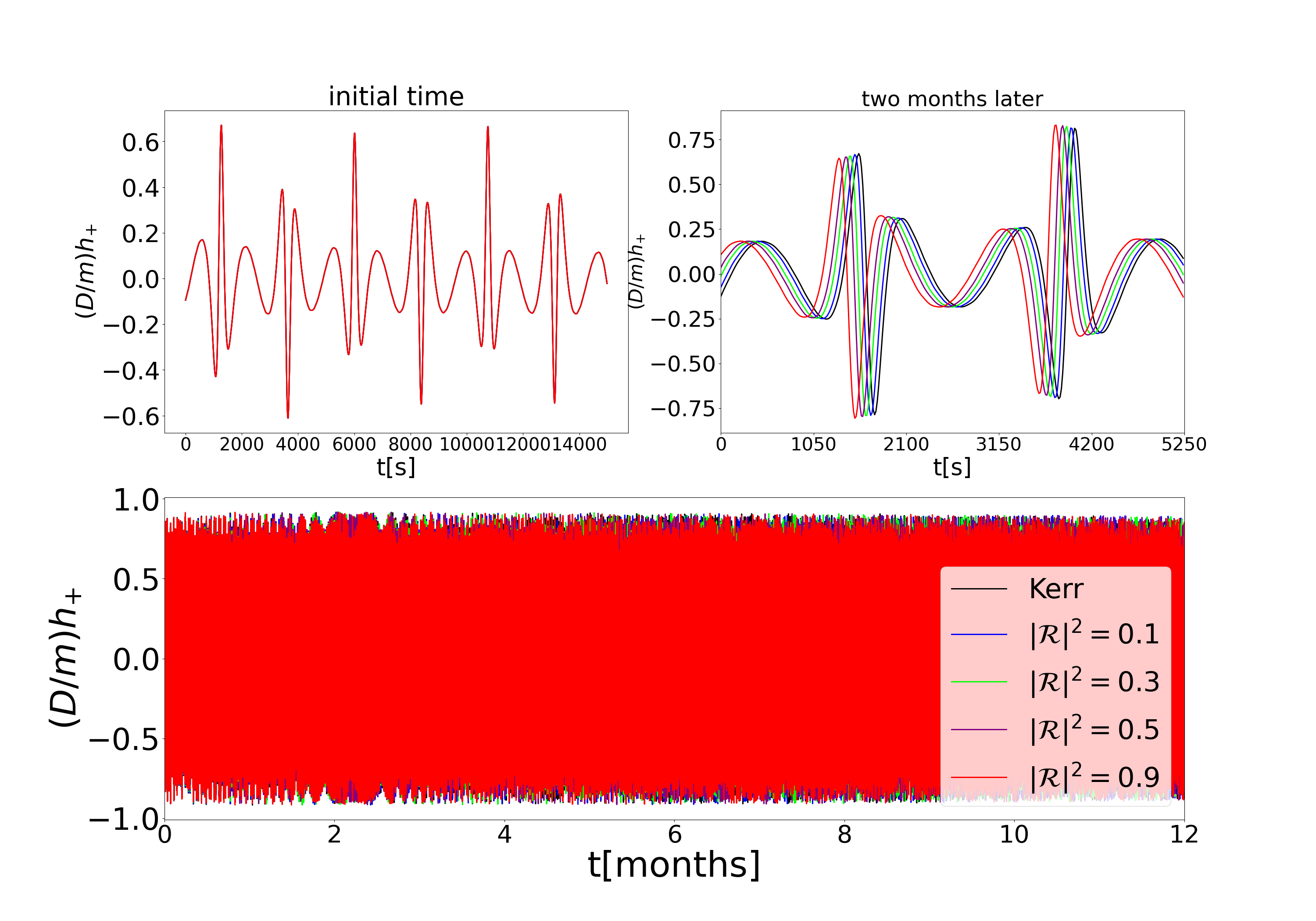}
 \caption{Comparison of the time domain EMRI waveforms from  the Kerr BH and ECO  with a reflectivity $|\mathcal{R}|^2=0.9$ and spin $a=0.9$, where the initial orbital parameters are $p_0=12, e_0=0.5, \theta_I=\pi/3$.}\label{wave:Kerr:NonKerr}
\end{figure*}

Firstly, we present a comparison of EMRI waveforms generated with the modified \texttt{FEW} model. In all the following plots, the massive object has a mass of $10^6 M_\odot$, a spin of $a=0.9$, a reflectivity of $|\mathcal{R}|^2=0.9$, the smaller object has a mass of $30 M_\odot$, and the distance of the EMRI is $D_L=1 \rm Gpc$.  Note that the subsequent figures retain  the same configuration.
As illustrated in Fig. \ref{wave:Kerr:NonKerr}, initially, the phases of the two EMRI waveforms closely coincide in the left panel of the top figure. However, as time progresses, specifically two months later, distinct disparities in the phases of the two EMRI waveforms become more pronounced in the right panel of the top figure.
The bottom of the figure displays the complete time domain waveform within one year.
Based on the left panel and right panel of upper part in the Fig. \ref{wave:Kerr:NonKerr}, one can find that the GW phases for the Kerr cases keep same with the phases for the EMRI system with the tidal heating parameters $|\mathcal{R}|^2\in\{0.1,0.3,0.5,0.9\}$, after about two months the phases differences between the Kerr case and the ECO' s case gradually increase over time.

\begin{figure*}[th]
 \centering
  \includegraphics[width=0.495\textwidth]{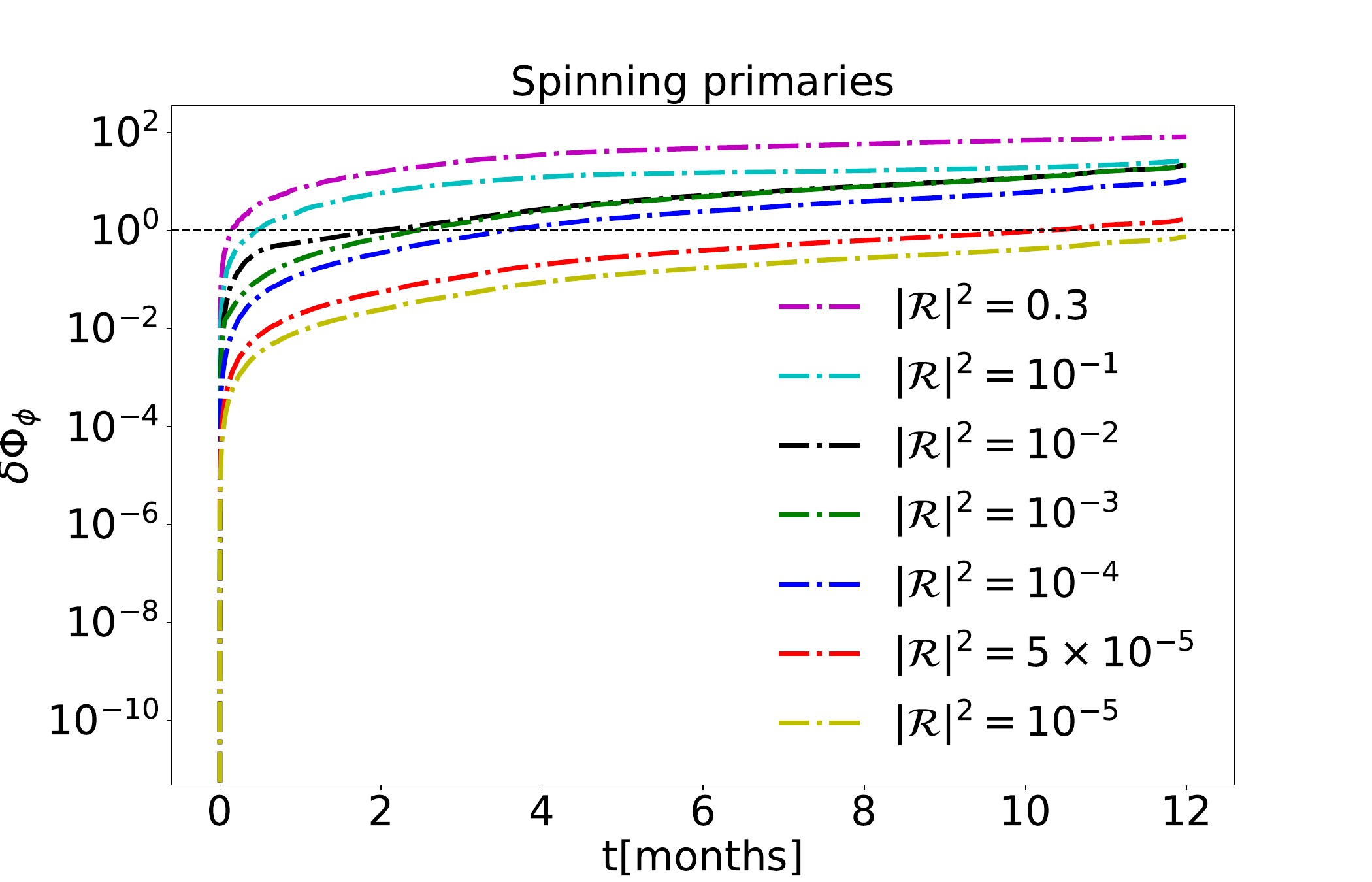}
  \includegraphics[width=0.495\textwidth]{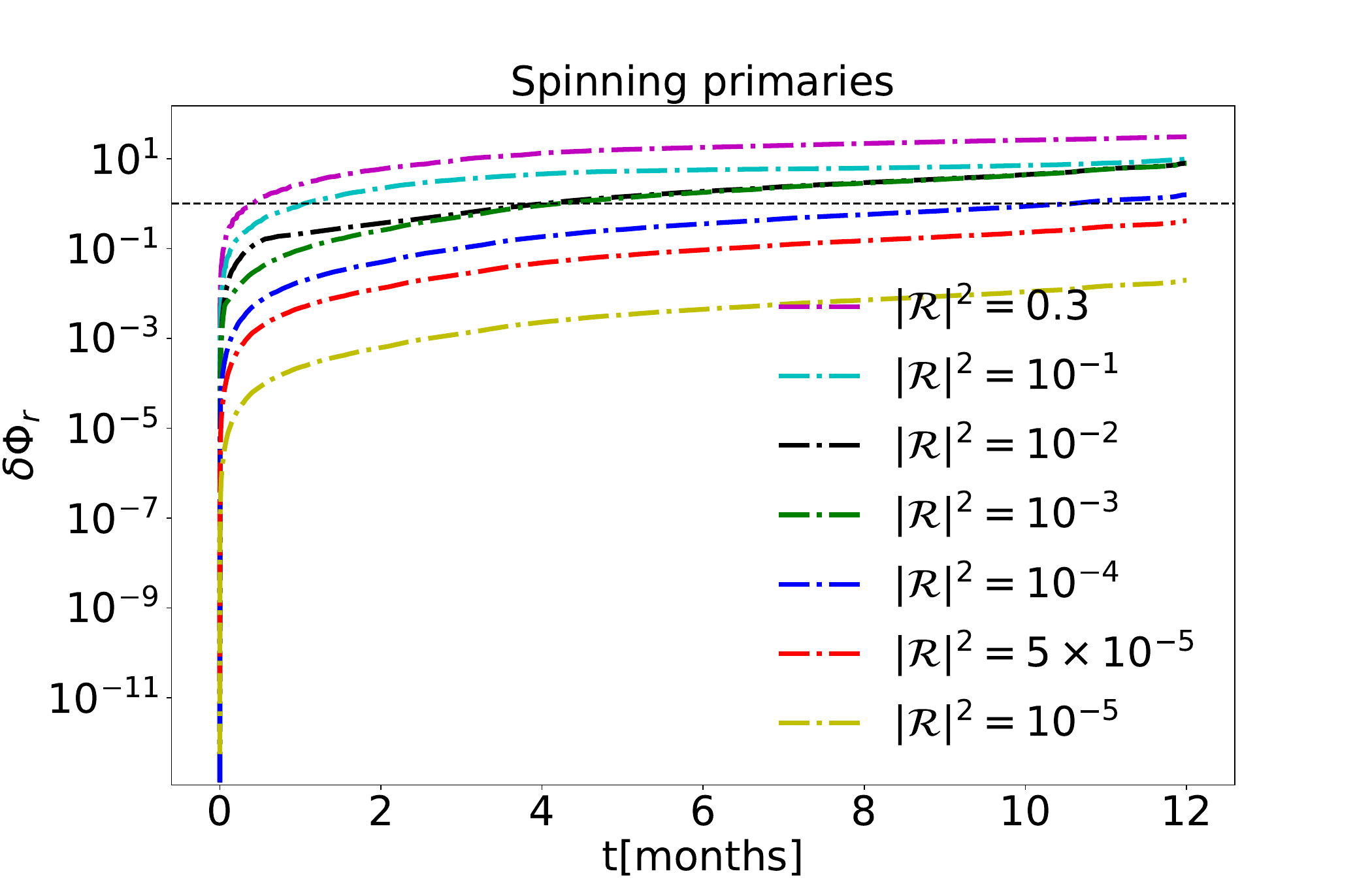}
  \includegraphics[width=0.495\textwidth]{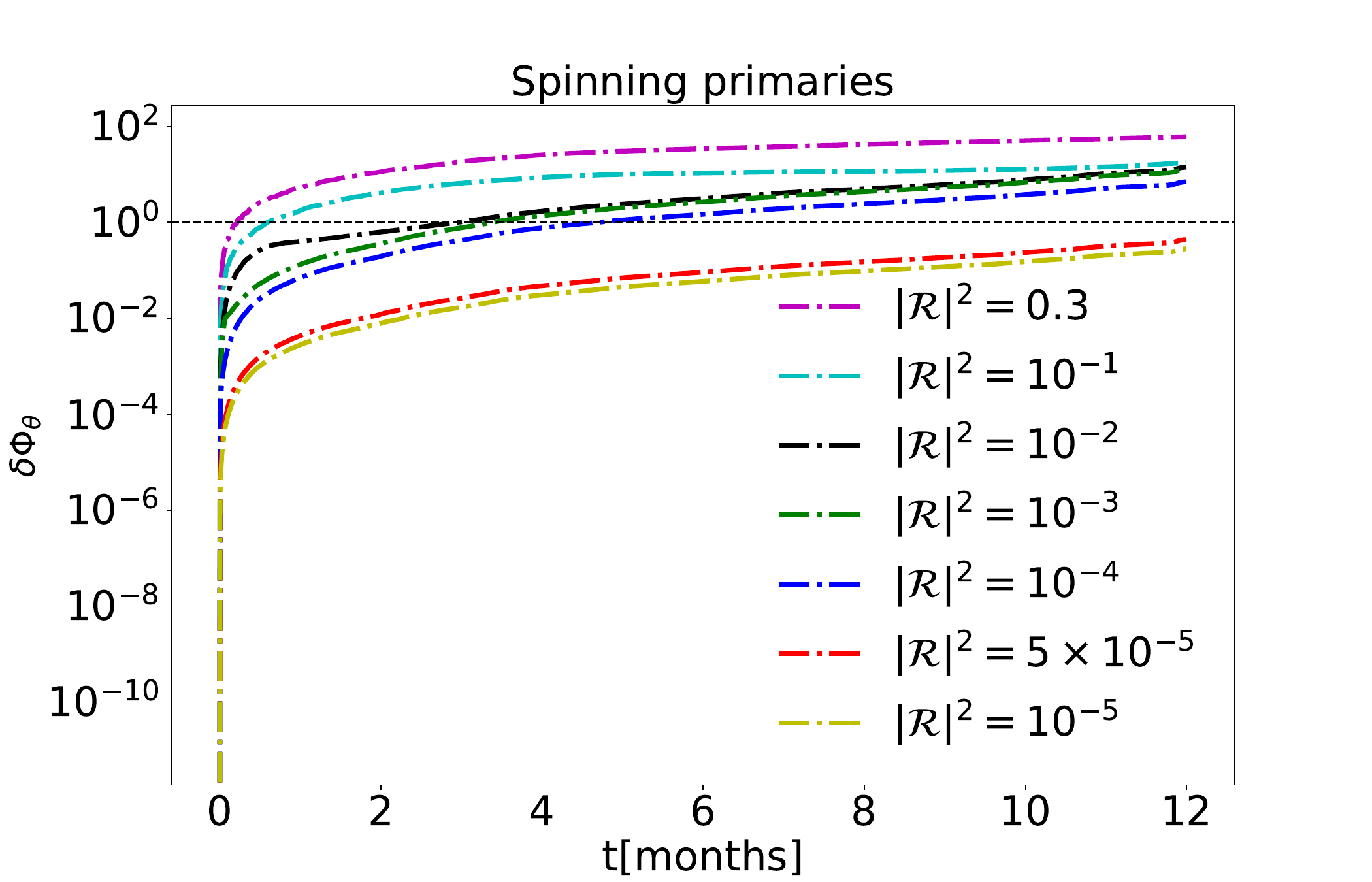}
 \caption{The azimuthal  dephasing $\delta \Phi_\phi$ (top left panel), radial dephasing $\delta \Phi_r$ (top right panel) and  polar dephasing $\delta \Phi_\theta$ of the EMRI waveforms as a function of observation time with various values of the reflectivity $|\mathcal{R}|^2 \in\{10^{-5},5\times 10^{-5},10^{-4},10^{-3},10^{-2},10^{-1},0.3\}$
 are depicted,  where the initial orbital  parameters are $p_0=12, e_0 = 0.5, \theta_I = \pi/3$.
  }\label{deltaphi:spinning}
\end{figure*}

The azimuthal, radial and polar dephasing of EMRI waveforms, as a function of observation time, is plotted in Fig. \ref{deltaphi:spinning}, where the initial orbital parameters are set the same as the ones in Fig. \ref{wave:Kerr:NonKerr}.
The black and dashed horizontal lines in Fig. \ref{deltaphi:spinning} represent the threshold values $\delta \Phi_i \sim 1 \rm rad$. This provides a very rough constraint on the tidal heating effect, and the shortcoming of this standard  has been  argued in Ref. \cite{Datta:2019epe}.
All the three panels for the dephasings of different coordinates are confirmed to be capable of exceeding  the threshold value when the suitable parameters are taken.
One can see that the accumulated dephasing varies from the fraction radians to $10^2$ radians as a whole, particularly for the azimuthal dephasing $\Omega_\phi$.
As argued in the Ref. \cite{Barsanti:2022ana,Gupta:2021cno},
the azimuthal dephasing plays the dominant role compared to the dephasings of the other coordinates.
According to Fig. \ref{deltaphi:spinning}, the radial and polar dephasings are both less than the azimuthal dephasing, which is consistent with the results in the Ref. \cite{Barsanti:2022ana,Gupta:2021cno}.
After considering the proper parameters, the accumulative effect of radial dephasing could be discernible.
In summary, the tidal heating effect can lead to a significant distinguishing criterion in the EMRI observations.

\begin{figure*}[th]
 \centering
 \includegraphics[width=0.4955\textwidth]{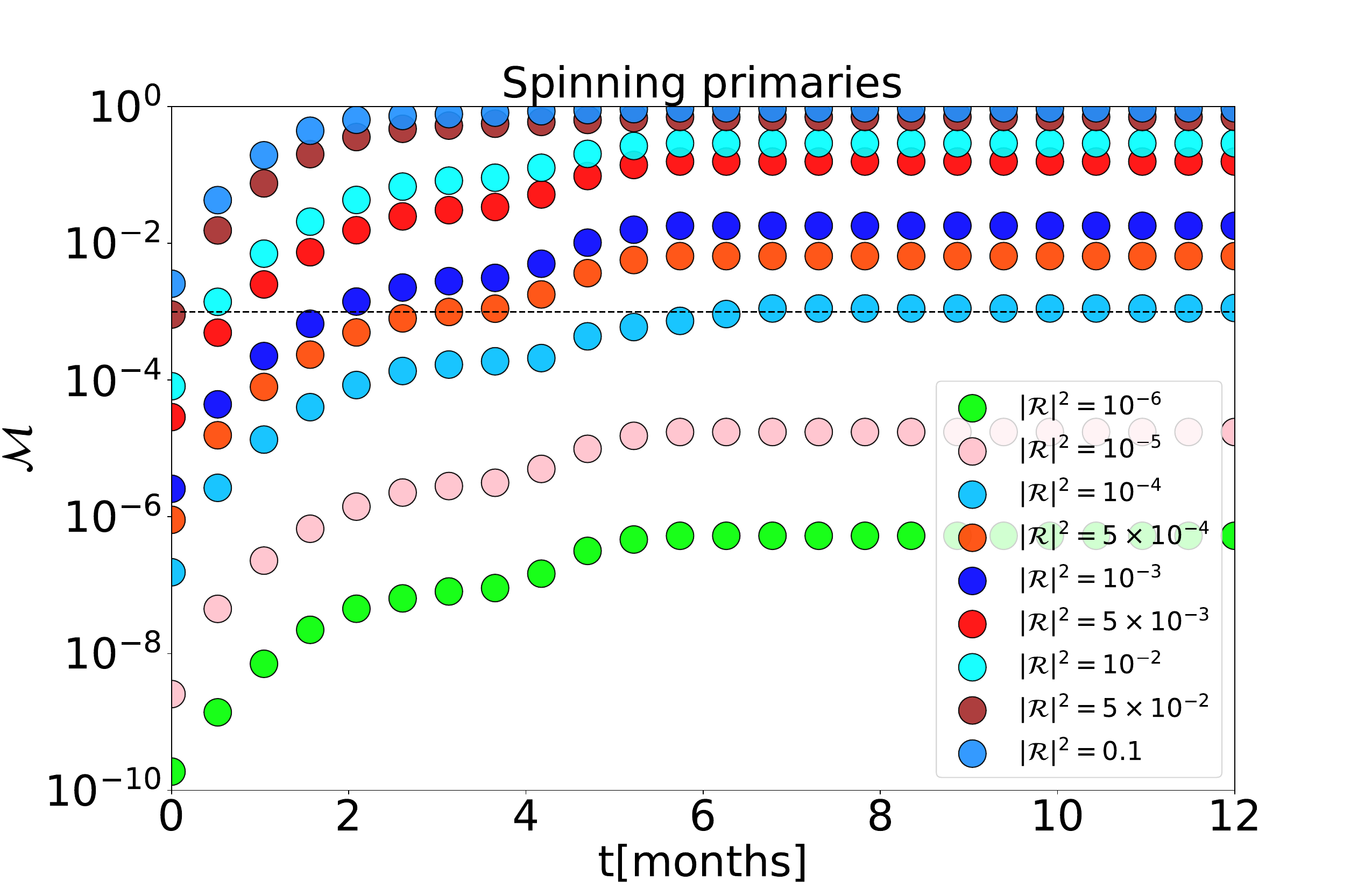}
  \includegraphics[width=0.4955\textwidth]{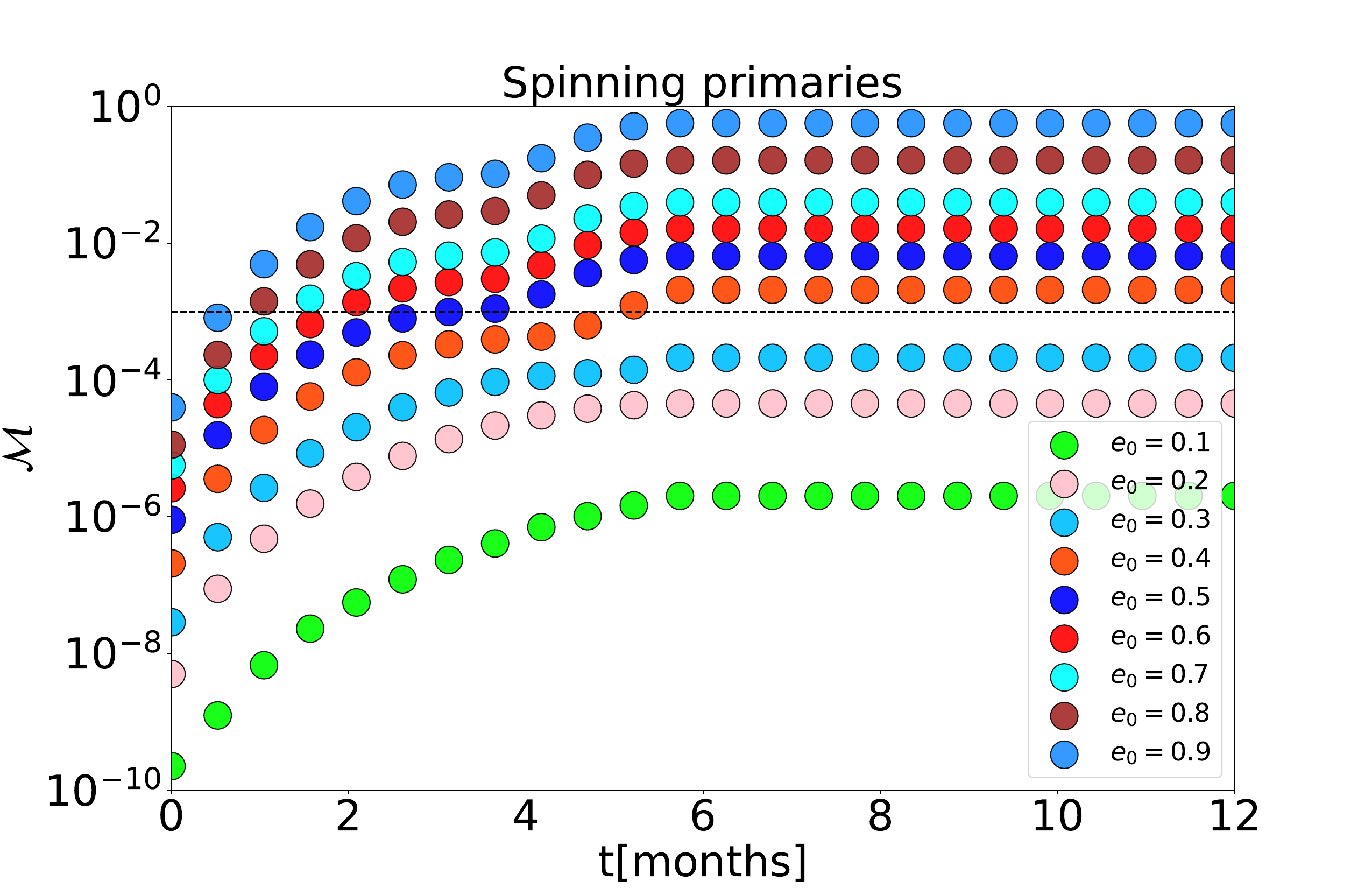}
 \caption{Mismatch as a function of observation time for EMRI waveforms from the ECO and the Kerr BH are depicted.
 The left panel shows the effects from various reflectivity parameters $|\mathcal{R}|^2 \in\{10^{-6},5\times10^{-5},10^{-4},5\times10^{-4},10^{-3},5\times10^{-3} , 10^{-2},0.05,0.1\}$ with  $e_0= 0.5$, and the right panel focus on the effects of orbital eccentricities $e_0\in\{0.1,0.2,0.3,0.4,0.5,0.6,0.7,0.8,0.9\}$ with $|\mathcal{R}|^2=5\times10^{-4}$  on the mismatch. The other parameters are set as $a=0.9, p_0=12, \theta_I = \pi/3$. }\label{Mismatch:spinning}
\end{figure*}

Though the aforementioned dephasing results could give the standard that identified the effect of tidal heating on EMRI waveforms,
we cannot claim that the minimum value of reflectivity can be distinguished with the LISA observation.
Thus, we perform the mismatch analysis using two kinds of waveforms from the EMRI systems where one is the Kerr waveform and the other is the ECO's.
In Fig. \ref{Mismatch:spinning}, the mismatches as a function of observation time for different values of the reflectivity and the initial orbital eccentricity are plotted. Note that the horizontal black dashed line denotes to the threshold value of mismatch can be discerned by LISA.
As a whole, after setting proper initial conditions, the mismatch can be larger than the threshold value, which implies that the tidal heating effect is distinguished in the EMRI observations.
According to mismatches results in Ref. \cite{Datta:2019epe},  EMRI waveforms from  the circular and equatorial orbits near the ECO with a reflectivity parameter $|\mathcal{R}|^2 = 10^{-4}$ can be just distinguished with thirteen months observation of LISA.
From the left panel of Fig. \ref{Mismatch:spinning}, however, the EMRI waveform corrected by tidal heating effect,  emitting from the eccentrical EMRI orbits, can be recognized for the initial orbital eccentric $e_0=0.5$ with about seven months observation of LISA.
As shown in the left panel, the mismatch increases with the reflectivity parameter $|\mathcal{R}|^2$.
For the right panel, the similar conclusions can be obtained if the initial orbital eccentricity is increasing.
For a given initial eccentricity $e_0=0.4$ and $|\mathcal{R}|^2=5\times 10^{-4}$, LISA can discern the tidal heating effect through a five-month observation period.

\begin{table*}[th]
\centering
\begin{tabular}{ccccccccc|ccc}
\multicolumn{11}{c}{Estimation uncertainty of EMRI parameters for the spinning massive object with $a = 0.9$}\\
\hline
\hline
$p_0$ & $e_0$ &$\Delta\ln M$  &$\Delta\ln m$  & $\Delta a$ & $\Delta e_0 $ &$\Delta p_0$  &$\Delta x_I$ &$\Delta(|\mathcal{R}|^2) $ & $ \Delta D_L/D_L$ & $\Delta \Omega_S$ & $\Delta \Omega_K$\\
\hline
$12$ & $0.4$  &$2.35\times10^{-5}$  &$1.23\times10^{-5}$  &$1.75\times10^{-6}$ & $1.39\times10^{-6}$
   &$2.18\times10^{-5}$   &$2.78\times10^{-6}$ &$1.82\times 10^{-6}$   &$1.82\times 10^{-2}$
 &$6.09\times 10^{-4}$   & $4.45$ \\
\hline
$12$ & $0.8$  &$1.43\times10^{-5}$  &$6.04\times10^{-7}$  &$8.28\times10^{-7}$ & $1.21\times10^{-6}$
   &$6.26\times10^{-4}$   &$1.43\times10^{-6}$   &$2.09\times 10^{-6}$   &$4.69\times 10^{-2}$
 &$5.82\times 10^{-4}$   & $2.75$ \\
\hline
\hline
\end{tabular}
\caption{Measurement errors of the EMRI source parameters including the reflectivity are listed, the other parameters are set as $\Phi_r^0=1.0$, $\Phi_\theta^0=1.0$, $\Phi_\phi^0=1.0$ and $|\mathcal{R}|^2=0.001$.  }\label{tab:fim:spin}
\end{table*}

Finally, we calculate the accuracy of source parameters, including reflectivity, using the FIM method. The EMRI waveforms are generated by the \texttt{FEW} model.
In Table \ref{tab:fim:spin}, we list the measurement errors on the source parameters with the LISA observations when the tidal heating effect is considered.
From the table, we find that the reflectivity parameter $|\mathcal{R}|^2$ of the ECO can be
constrained within a fractional error of $~10^{-6}$ via the observations of LISA.
The measurement errors of the full waveform parameters is slightly decreased with the initial orbital eccentricity.

\begin{figure*}
    \centering
    \includegraphics[width=2.1\columnwidth]{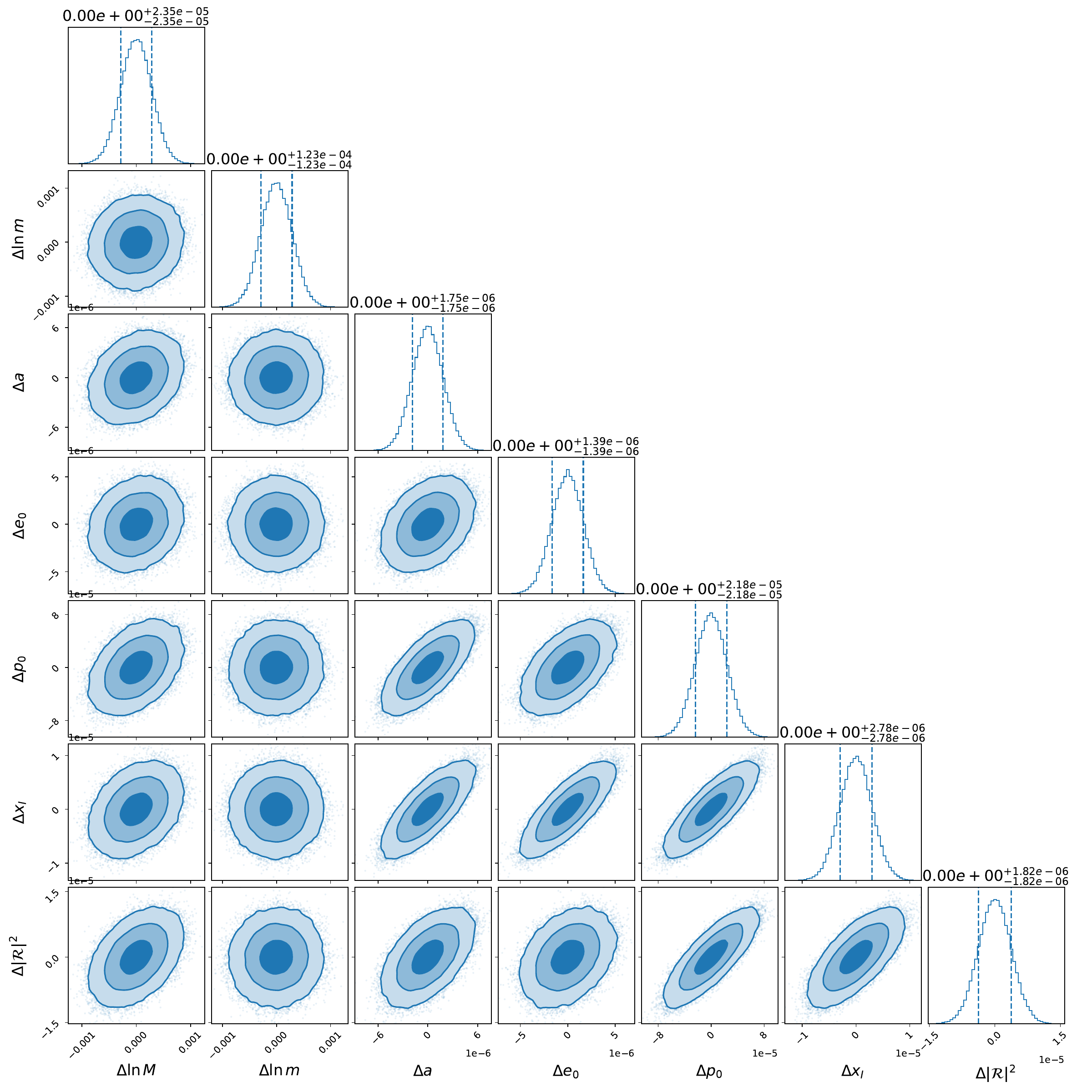}
    \caption{The corner plot for the probability distribution for the mass of secondary body, mass of primary body, spin of primary body, initial orbital eccentricity, initial semi-latus rectum, initial orbital inclination and reflectivity $(M,m,a=0.9,e_0=0.4,p_0=12,x_I, |\mathcal{R}|^2)$, which is inferred from one-year observation of EMRI. Vertical lines show the $1\sigma$ interval for each source parameter.
    The contours correspond to $68\%$, $95\%$ and $99\%$ probability confidence intervals}
    \label{fig4:corner}
\end{figure*}

In most papers on EMRI data analysis and parameter estimation  \cite{Babak:2009ua,Wang:2012xh,Ye:2024bku}, the focus is primarily on the correlations among intrinsic parameters  \cite{Katz:2021yft,Zhang:2022rfr,Zhang:2023vok}. Therefore, in the following subsection, we only plot the correlations between the constraint reflectivity and the other intrinsic parameters.
Fig. \ref{fig4:corner} shows the probability distribution for the mass of secondary body, mass of primary body, spin of primary body, initial orbital eccentricity, initial semi-latus rectum, initial orbital inclination and reflectivity $(M,m,a,e_0,p_0,x_I, |\mathcal{R}|^2)$, the results are plotted using the off-diagonal elements of the variance-covariance matrix. One can find that the bound on the reflectivity is closely associated with the parameters $(M,a,e_0,p_0,x_I)$, and the correlation between the reflection $|\mathcal{R}|^2$ and secondary body mass
 $m$ is relative weaker.

\subsection{Non-spinning massive objects}\label{result:non-spinning}
In the subsection we present similar results by employing EMRI waveforms from non-spinning ECOs.
The EMRI trajectories modified by the ECOs are computed utilizing the analytic fluxes in the Schwarzschild spacetime \cite{Munna:2020juq,Munna:2023vds}.

\begin{figure*}[th]
 \centering
  \includegraphics[width=0.495\textwidth]{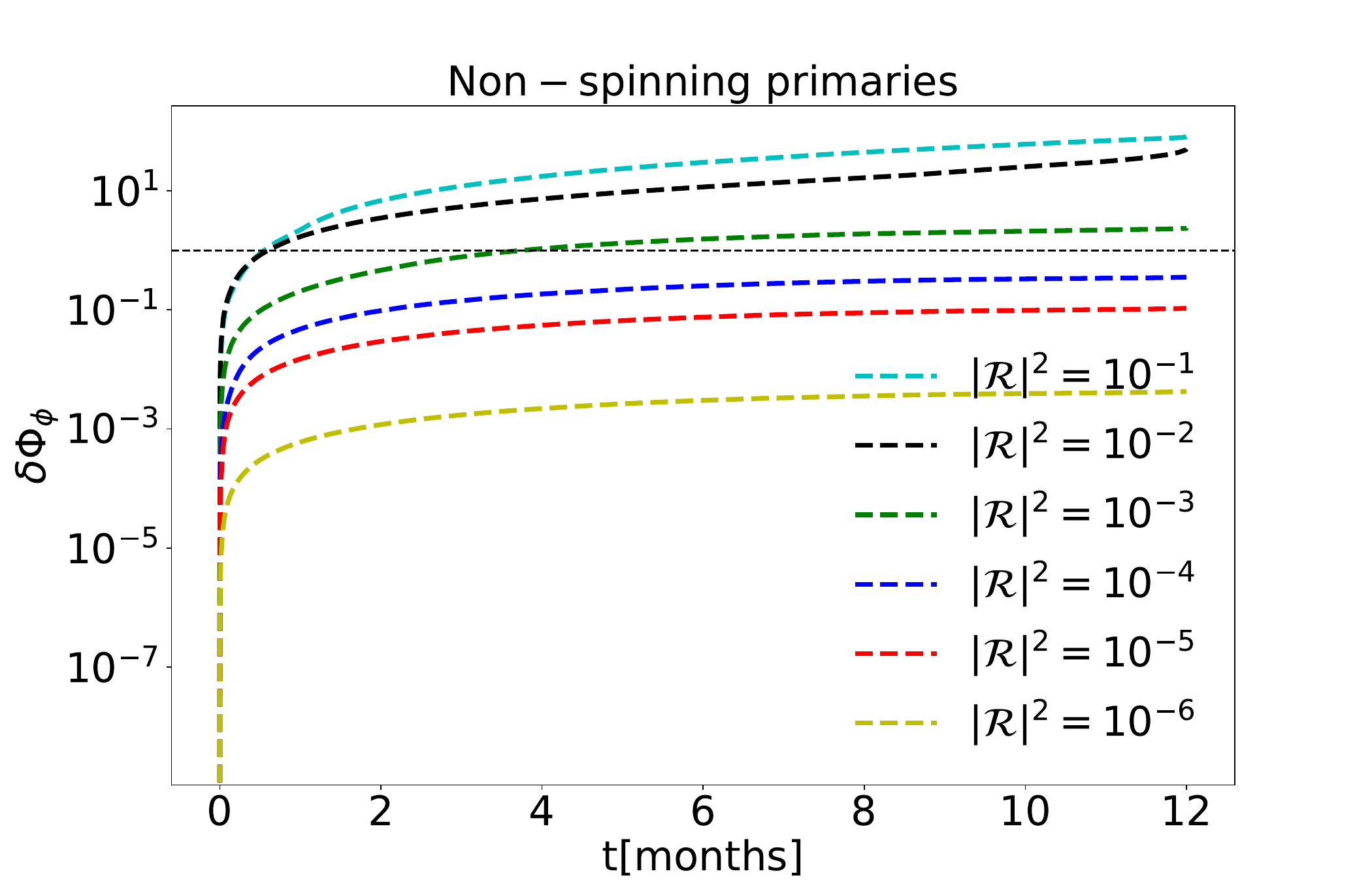}
 \caption{The azimuthal  dephasing $\delta \Phi_\phi$ as a function of the observation time is plotted.  The parameters are taken as  $p_0=12, e_0 = 0.5, \theta_I = \pi/3, |\mathcal{R}|^2 \in(10^{-6}, 10^{-5},10^{-4},10^{-3},10^{-2},10^{-1})$.
  }\label{deltaphi:non-spinning}
\end{figure*}

Since the azimuthal, radial and polar dephasings exhibit the similar behaviors, except for the difference in magnitudes, here we only show the azimuthal dephasing as a function of observation time in Fig. \ref{deltaphi:non-spinning}. The horizontal black dashed line denotes the threshold value that can be distinguished by LISA. For the EMRI waveforms from the ECO with $|\mathcal{R}|^2=10^{-3}$ and the initial eccentricity $e_0=0.5$, the tidal heating effect can be discerned with the observation of LISA of four months.

\begin{figure*}[th]
 \centering
 \includegraphics[width=0.4955\textwidth]{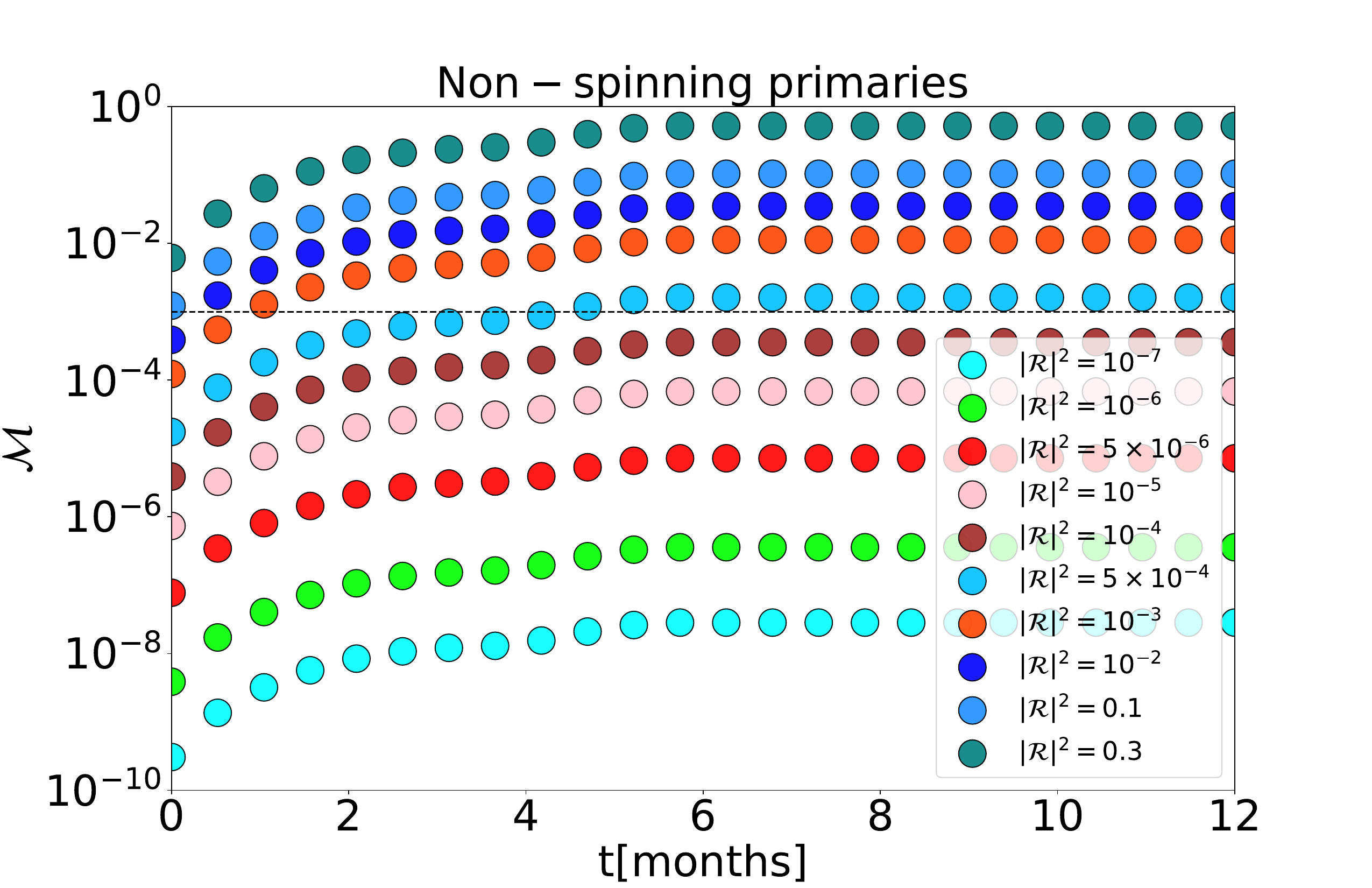}
  \includegraphics[width=0.4955\textwidth]{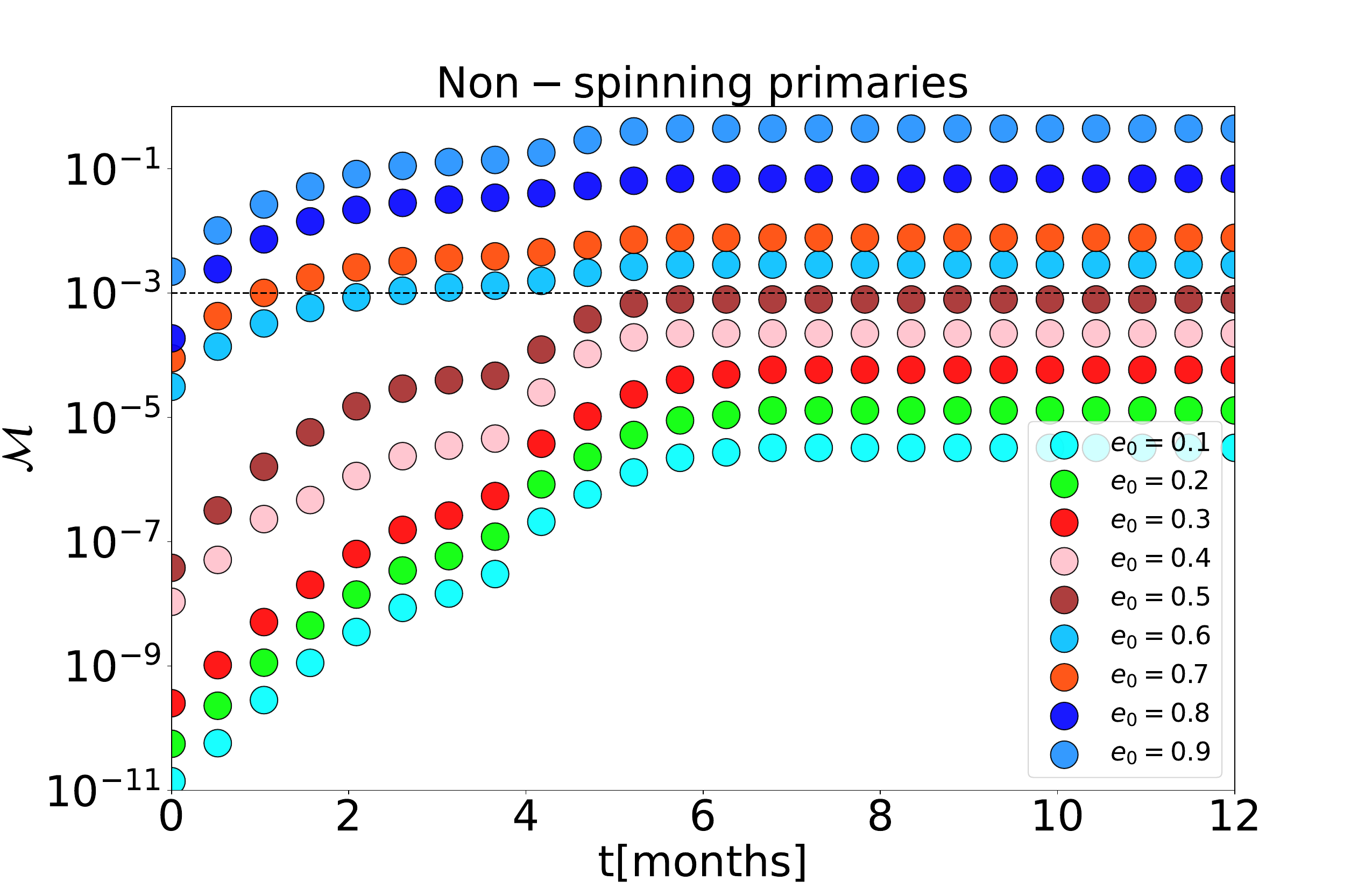}
 \caption{Mismatch as a function of observation time for EMRI waveforms from ECO and Schwarzschild BH is plotted, where the other parameters are set as $p_0=12, \theta_I = \pi/3$.
 The left panel shows the effects from various values of the reflectivity parameter with $e_0= 0.7$, and the right panel focuses the effects from various values of the orbital eccentricity with$ |\mathcal{R}|^2=10^{-3}$.  }\label{Mismatch:nonspinning}
\end{figure*}

Then we also analyze the mismatch of the two EMRI waveforms from the Schwarzschild BH and the non-spinning ECO cases in Fig. \ref{Mismatch:nonspinning}.
From the  left panels of Fig. \ref{Mismatch:spinning} and Fig. \ref{Mismatch:nonspinning},
one can see that with the same sources parameters, the mismatch in the non-spinning case is slightly smaller than that in the spinning case.
From the right panel of Fig. \ref{Mismatch:nonspinning}, as a whole, the mismatch grows  with the initial eccentricity. For an initial eccentricity $e_0=0.5$, LISA can not discern the difference of the tidal heating effect with one year observation for the non-spinning primaries, however, the observation of four months can be distinguished for the spinning primaries in the right panel of Fig. \ref{Mismatch:spinning}.

\begin{table*}[th]
\centering
\begin{tabular}{ccccccc|ccc}
\multicolumn{10}{c}{Estimation uncertainty of EMRI parameters for the non-spinning massive object}\\
\hline
\hline
$p_0$ & $e_0$ &$\Delta \ln M$  &$\Delta \ln m$  & $\Delta e_0 $ &$\Delta p_0$  &$\Delta|\mathcal{R}|^2 $ & $ \Delta D_L/D_L$ & $\Delta \Omega_S$ & $\Delta \Omega_K$\\
\hline
$12$ & $0.4$  &$1.19\times10^{-4}$  &$2.44\times10^{-3}$   & $1.61\times10^{-3}$
   &$9.18\times10^{-4}$   &$1.03\times 10^{-3}$   &$8.67\times 10^{-2}$
 &$7.59\times 10^{-3}$   & $7.86$ \\
\hline
$12$ & $0.8$  &$7.51\times10^{-5}$  &$8.28\times10^{-4}$  & $1.45\times10^{-3}$
   &$8.71\times10^{-4}$    &$5.42\times 10^{-4}$   &$4.65\times 10^{-2}$
 &$5.54\times 10^{-3}$   & $5.62$ \\
\hline
\hline
\end{tabular}
\caption{Measurement errors of the full EMRI source parameters are listed, the other parameters are set as $\Phi_r^0=1.0$, $\Phi_\phi^0=1.0$ and $|\mathcal{R}|^2=0.001$.  }\label{tab:fim:nonspin}
\end{table*}

Finally, we also study the constraint on the reflectivity of the non-spinning ECO using the FIM method.
The results of the estimation accuracy  of the EMRI source parameters are listed in Table \ref{tab:fim:nonspin}.
It is evident that the constraint on reflectivity for the non-spinning ECO can reach levels on the order of $\sim10^{-4}$, mirroring the situation in the spinning ECO case.
Furthermore, the measurement accuracy of the reflectivity  increases slightly with a larger initial eccentricity.
\begin{figure*}
    \centering
    \includegraphics[width=2.1\columnwidth]{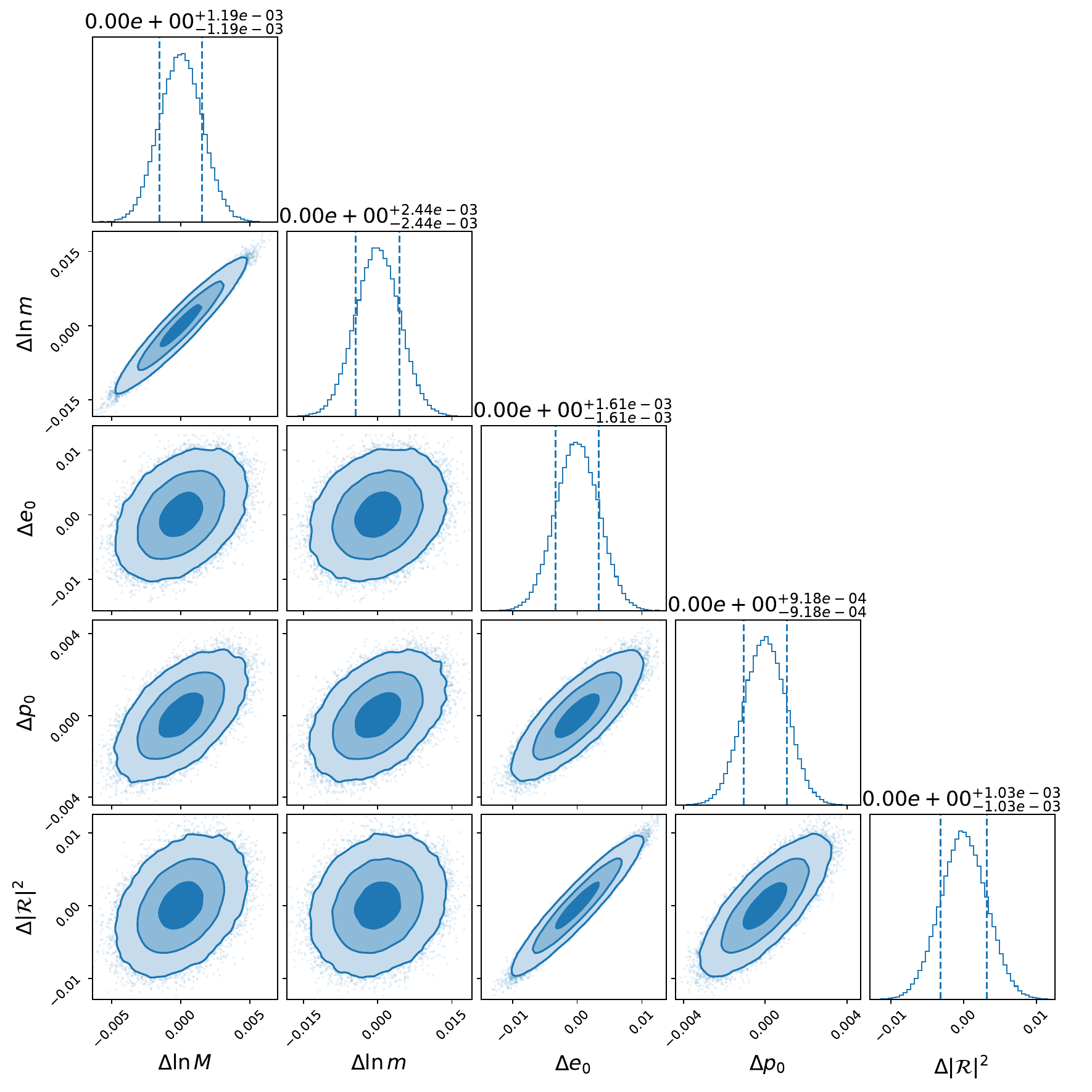}
    \caption{The corner plot for the probability distribution for the mass of secondary body, mass of primary body, initial orbital eccentricity, initial semi-latus rectum and reflectivity $(M,m,e_0=0.4,p_0=12, |\mathcal{R}|^2)$, which is inferred from one-year observation of EMRI. Vertical lines show the $1\sigma$ interval for each source parameter.
    The contours correspond to $68\%$, $95\%$ and $99\%$ probability confidence intervals}
    \label{fig7:corner}
\end{figure*}

In the end, we also argue the correlations of constraint reflectivity among the other intrinsic parameters in the non-spinning primary case.
We display the probability distribution for the mass of secondary body, mass of primary body, initial orbital eccentricity, initial semi-latus rectum  and reflectivity $(M,m,e_0,p_0, |\mathcal{R}|^2)$ in Fig. \ref{fig7:corner}. One can find that the bound on the reflectivity is closely associated with the parameters $(e_0,p_0)$, and the correlation between the reflectivity $|\mathcal{R}|^2$ and celestial bodies $(m,M)$ is relative weaker.

\section{Conclusion}\label{Conclusion}
A key and interesting goal of the EMRI detections is to probe the nature of dark compact objects \cite{Cardoso:2019rvt}.
Recent studies have developed proposals to distinguish massive horizonless objects based on considerations of tidal heating
\cite{Datta:2019euh,Datta:2019epe}, tidal deformability \cite{Pani:2019cyc,DeLuca:2022xlz,Zi:2023pvl} and  area quantization \cite{Agullo:2020hxe,Datta:2021row}.
In this paper we analyzed the difference of waveforms from the generic EMRIs around the ECOs, then calculated the constraint on the reflectivity parameter with the FIM method.

Utilizing precise analytic flux formulas derived for Kerr \cite{Sago:2015rpa} and Schwarzschild BHs \cite{Munna:2020juq,Munna:2023vds}, we adapted these formulas in the vicinity of the ECO surface by incorporating a reflectivity parameter, employing the methodology outlined in Ref. \cite{Datta:2019epe}.
Injecting the modified fluxes into the \texttt{FEW} model, we computed the EMRI waveforms of the ECOs. Then we analyzed the differences of EMRI waveforms from the ECO and
the Kerr  or Schwarzschild BH spacetime  by computing the dephasing and mismatch. Our results indicate that one year  observation of EMRI by LISA can distinguish the tidal heating corrected signals with
a reflectivity as smaller as $|\mathcal{R}|^2\sim10^{-6}$ for the spinning ECO, however,
the EMRI signal from the non-spinning ECO can constrain the reflectivity parameter at the level of $|\mathcal{R}|^2\sim5\times 10^{-4}$.
Finally, we assessed the capability of LISA to constrain the tidal heating effect using the FIM method.
The measurement accuracy of the reflectivity of the ECOs through LISA observations can achieve a precision on the order of $\mathcal{O}(10^{-6})$. The constraint level is expected to be further enhanced with the incorporation of the orbital eccentricity.
According to Figs. \ref{fig4:corner} and \ref{fig7:corner},
Note that the reflectivity $|\mathcal{R}|^2$ and the other parameters $(M,a,e_0,p_0,x_I)$
are correlated for the rotating ECO case, the correlations between reflectivity $|\mathcal{R}|^2$ and the other parameters $(M,e_0,p_0)$ are positive for the non-spinning ECO case.
In summary, the reflectivity of the spinning ECO could be more rigorously constrained using eccentric EMRI signals compared to the non-spinning ECO case.

In this paper we only used the modified fluxes at the leading order in the mass ratio to evolve the trajectories in the \texttt{FEW} model, with the conservative part of self-fore being ignored \cite{Poisson:2003nc}.
Given that the long time scale inspiral dynamics of EMRIs, the conservative part of self-fore
also plays an important role in the waveform generation and data analysis.
Therefore, to study the effect of tidal heating the full contribution of the self-fore
should be included in the future.

\section*{Acknowledgments}
The work is in part supported by NSFC Grant
No.12205104, ``the Fundamental Research Funds for the Central Universities'' with Grant No.  2023ZYGXZR079, the Guangzhou Science and Technology Project with Grant No. 2023A04J0651 and the startup funding of South China University of Technology.
T. Zi is  funded by the China Postdoctoral Science Foundation with  Grant
No. 2023M731137 and the National Natural Science Foundation
of China with Grants No. 12347140.
T.Zi thanks Chao Zhang for very helpful discussions on the corner plots.


\begin{thebibliography}{83}%
\makeatletter
\providecommand \@ifxundefined [1]{%
 \@ifx{#1\undefined}
}%
\providecommand \@ifnum [1]{%
 \ifnum #1\expandafter \@firstoftwo
 \else \expandafter \@secondoftwo
 \fi
}%
\providecommand \@ifx [1]{%
 \ifx #1\expandafter \@firstoftwo
 \else \expandafter \@secondoftwo
 \fi
}%
\providecommand \natexlab [1]{#1}%
\providecommand \enquote  [1]{``#1''}%
\providecommand \bibnamefont  [1]{#1}%
\providecommand \bibfnamefont [1]{#1}%
\providecommand \citenamefont [1]{#1}%
\providecommand \href@noop [0]{\@secondoftwo}%
\providecommand \href [0]{\begingroup \@sanitize@url \@href}%
\providecommand \@href[1]{\@@startlink{#1}\@@href}%
\providecommand \@@href[1]{\endgroup#1\@@endlink}%
\providecommand \@sanitize@url [0]{\catcode `\\12\catcode `\$12\catcode
  `\&12\catcode `\#12\catcode `\^12\catcode `\_12\catcode `\%12\relax}%
\providecommand \@@startlink[1]{}%
\providecommand \@@endlink[0]{}%
\providecommand \url  [0]{\begingroup\@sanitize@url \@url }%
\providecommand \@url [1]{\endgroup\@href {#1}{\urlprefix }}%
\providecommand \urlprefix  [0]{URL }%
\providecommand \Eprint [0]{\href }%
\providecommand \doibase [0]{http://dx.doi.org/}%
\providecommand \selectlanguage [0]{\@gobble}%
\providecommand \bibinfo  [0]{\@secondoftwo}%
\providecommand \bibfield  [0]{\@secondoftwo}%
\providecommand \translation [1]{[#1]}%
\providecommand \BibitemOpen [0]{}%
\providecommand \bibitemStop [0]{}%
\providecommand \bibitemNoStop [0]{.\EOS\space}%
\providecommand \EOS [0]{\spacefactor3000\relax}%
\providecommand \BibitemShut  [1]{\csname bibitem#1\endcsname}%
\let\auto@bib@innerbib\@empty
\bibitem [{\citenamefont {Hawking}(1976)}]{Hawking:1976ra}%
  \BibitemOpen
  \bibfield  {author} {\bibinfo {author} {\bibfnamefont {S.~W.}\ \bibnamefont
  {Hawking}},\ }\href {\doibase 10.1103/PhysRevD.14.2460} {\bibfield  {journal}
  {\bibinfo  {journal} {Phys. Rev. D}\ }\textbf {\bibinfo {volume} {14}},\
  \bibinfo {pages} {2460} (\bibinfo {year} {1976})}\BibitemShut {NoStop}%
\bibitem [{\citenamefont {Bianchi}\ \emph {et~al.}(2021)\citenamefont
  {Bianchi}, \citenamefont {Consoli}, \citenamefont {Grillo}, \citenamefont
  {Morales}, \citenamefont {Pani},\ and\ \citenamefont
  {Raposo}}]{Bianchi:2020miz}%
  \BibitemOpen
  \bibfield  {author} {\bibinfo {author} {\bibfnamefont {M.}~\bibnamefont
  {Bianchi}}, \bibinfo {author} {\bibfnamefont {D.}~\bibnamefont {Consoli}},
  \bibinfo {author} {\bibfnamefont {A.}~\bibnamefont {Grillo}}, \bibinfo
  {author} {\bibfnamefont {J.~F.}\ \bibnamefont {Morales}}, \bibinfo {author}
  {\bibfnamefont {P.}~\bibnamefont {Pani}}, \ and\ \bibinfo {author}
  {\bibfnamefont {G.}~\bibnamefont {Raposo}},\ }\href {\doibase
  10.1007/JHEP01(2021)003} {\bibfield  {journal} {\bibinfo  {journal} {JHEP}\
  }\textbf {\bibinfo {volume} {01}},\ \bibinfo {pages} {003} (\bibinfo {year}
  {2021})},\ \Eprint {http://arxiv.org/abs/2008.01445} {arXiv:2008.01445
  [hep-th]} \BibitemShut {NoStop}%
\bibitem [{\citenamefont {Giddings}(2017)}]{Giddings:2017jts}%
  \BibitemOpen
  \bibfield  {author} {\bibinfo {author} {\bibfnamefont {S.~B.}\ \bibnamefont
  {Giddings}},\ }\href {\doibase 10.1038/s41550-017-0067} {\bibfield  {journal}
  {\bibinfo  {journal} {Nature Astron.}\ }\textbf {\bibinfo {volume} {1}},\
  \bibinfo {pages} {0067} (\bibinfo {year} {2017})},\ \Eprint
  {http://arxiv.org/abs/1703.03387} {arXiv:1703.03387 [gr-qc]} \BibitemShut
  {NoStop}%
\bibitem [{\citenamefont {Maselli}\ \emph {et~al.}(2018)\citenamefont
  {Maselli}, \citenamefont {Pani}, \citenamefont {Cardoso}, \citenamefont
  {Abdelsalhin}, \citenamefont {Gualtieri},\ and\ \citenamefont
  {Ferrari}}]{Maselli:2017cmm}%
  \BibitemOpen
  \bibfield  {author} {\bibinfo {author} {\bibfnamefont {A.}~\bibnamefont
  {Maselli}}, \bibinfo {author} {\bibfnamefont {P.}~\bibnamefont {Pani}},
  \bibinfo {author} {\bibfnamefont {V.}~\bibnamefont {Cardoso}}, \bibinfo
  {author} {\bibfnamefont {T.}~\bibnamefont {Abdelsalhin}}, \bibinfo {author}
  {\bibfnamefont {L.}~\bibnamefont {Gualtieri}}, \ and\ \bibinfo {author}
  {\bibfnamefont {V.}~\bibnamefont {Ferrari}},\ }\href {\doibase
  10.1103/PhysRevLett.120.081101} {\bibfield  {journal} {\bibinfo  {journal}
  {Phys. Rev. Lett.}\ }\textbf {\bibinfo {volume} {120}},\ \bibinfo {pages}
  {081101} (\bibinfo {year} {2018})},\ \Eprint
  {http://arxiv.org/abs/1703.10612} {arXiv:1703.10612 [gr-qc]} \BibitemShut
  {NoStop}%
\bibitem [{\citenamefont {Agullo}\ \emph {et~al.}(2021)\citenamefont {Agullo},
  \citenamefont {Cardoso}, \citenamefont {Rio}, \citenamefont {Maggiore},\ and\
  \citenamefont {Pullin}}]{Agullo:2020hxe}%
  \BibitemOpen
  \bibfield  {author} {\bibinfo {author} {\bibfnamefont {I.}~\bibnamefont
  {Agullo}}, \bibinfo {author} {\bibfnamefont {V.}~\bibnamefont {Cardoso}},
  \bibinfo {author} {\bibfnamefont {A.~D.}\ \bibnamefont {Rio}}, \bibinfo
  {author} {\bibfnamefont {M.}~\bibnamefont {Maggiore}}, \ and\ \bibinfo
  {author} {\bibfnamefont {J.}~\bibnamefont {Pullin}},\ }\href {\doibase
  10.1103/PhysRevLett.126.041302} {\bibfield  {journal} {\bibinfo  {journal}
  {Phys. Rev. Lett.}\ }\textbf {\bibinfo {volume} {126}},\ \bibinfo {pages}
  {041302} (\bibinfo {year} {2021})},\ \Eprint
  {http://arxiv.org/abs/2007.13761} {arXiv:2007.13761 [gr-qc]} \BibitemShut
  {NoStop}%
\bibitem [{\citenamefont {Bianchi}\ \emph {et~al.}(2020)\citenamefont
  {Bianchi}, \citenamefont {Consoli}, \citenamefont {Grillo}, \citenamefont
  {Morales}, \citenamefont {Pani},\ and\ \citenamefont
  {Raposo}}]{Bianchi:2020bxa}%
  \BibitemOpen
  \bibfield  {author} {\bibinfo {author} {\bibfnamefont {M.}~\bibnamefont
  {Bianchi}}, \bibinfo {author} {\bibfnamefont {D.}~\bibnamefont {Consoli}},
  \bibinfo {author} {\bibfnamefont {A.}~\bibnamefont {Grillo}}, \bibinfo
  {author} {\bibfnamefont {J.~F.}\ \bibnamefont {Morales}}, \bibinfo {author}
  {\bibfnamefont {P.}~\bibnamefont {Pani}}, \ and\ \bibinfo {author}
  {\bibfnamefont {G.}~\bibnamefont {Raposo}},\ }\href {\doibase
  10.1103/PhysRevLett.125.221601} {\bibfield  {journal} {\bibinfo  {journal}
  {Phys. Rev. Lett.}\ }\textbf {\bibinfo {volume} {125}},\ \bibinfo {pages}
  {221601} (\bibinfo {year} {2020})},\ \Eprint
  {http://arxiv.org/abs/2007.01743} {arXiv:2007.01743 [hep-th]} \BibitemShut
  {NoStop}%
\bibitem [{\citenamefont {Mazur}\ and\ \citenamefont
  {Mottola}(2004)}]{Mazur:2004fk}%
  \BibitemOpen
  \bibfield  {author} {\bibinfo {author} {\bibfnamefont {P.~O.}\ \bibnamefont
  {Mazur}}\ and\ \bibinfo {author} {\bibfnamefont {E.}~\bibnamefont
  {Mottola}},\ }\href {\doibase 10.1073/pnas.0402717101} {\bibfield  {journal}
  {\bibinfo  {journal} {Proc. Nat. Acad. Sci.}\ }\textbf {\bibinfo {volume}
  {101}},\ \bibinfo {pages} {9545} (\bibinfo {year} {2004})},\ \Eprint
  {http://arxiv.org/abs/gr-qc/0407075} {arXiv:gr-qc/0407075} \BibitemShut
  {NoStop}%
\bibitem [{\citenamefont {Cardoso}\ and\ \citenamefont
  {Pani}(2019)}]{Cardoso:2019rvt}%
  \BibitemOpen
  \bibfield  {author} {\bibinfo {author} {\bibfnamefont {V.}~\bibnamefont
  {Cardoso}}\ and\ \bibinfo {author} {\bibfnamefont {P.}~\bibnamefont {Pani}},\
  }\href {\doibase 10.1007/s41114-019-0020-4} {\bibfield  {journal} {\bibinfo
  {journal} {Living Rev. Rel.}\ }\textbf {\bibinfo {volume} {22}},\ \bibinfo
  {pages} {4} (\bibinfo {year} {2019})},\ \Eprint
  {http://arxiv.org/abs/1904.05363} {arXiv:1904.05363 [gr-qc]} \BibitemShut
  {NoStop}%
\bibitem [{\citenamefont {Abbott}\ \emph
  {et~al.}(2019{\natexlab{a}})\citenamefont {Abbott} \emph
  {et~al.}}]{LIGOScientific:2018dkp}%
  \BibitemOpen
  \bibfield  {author} {\bibinfo {author} {\bibfnamefont {B.~P.}\ \bibnamefont
  {Abbott}} \emph {et~al.} (\bibinfo {collaboration} {LIGO Scientific,
  Virgo}),\ }\href {\doibase 10.1103/PhysRevLett.123.011102} {\bibfield
  {journal} {\bibinfo  {journal} {Phys. Rev. Lett.}\ }\textbf {\bibinfo
  {volume} {123}},\ \bibinfo {pages} {011102} (\bibinfo {year}
  {2019}{\natexlab{a}})},\ \Eprint {http://arxiv.org/abs/1811.00364}
  {arXiv:1811.00364 [gr-qc]} \BibitemShut {NoStop}%
\bibitem [{\citenamefont {Abbott}\ \emph
  {et~al.}(2021{\natexlab{a}})\citenamefont {Abbott} \emph
  {et~al.}}]{LIGOScientific:2020tif}%
  \BibitemOpen
  \bibfield  {author} {\bibinfo {author} {\bibfnamefont {R.}~\bibnamefont
  {Abbott}} \emph {et~al.} (\bibinfo {collaboration} {LIGO Scientific,
  Virgo}),\ }\href {\doibase 10.1103/PhysRevD.103.122002} {\bibfield  {journal}
  {\bibinfo  {journal} {Phys. Rev. D}\ }\textbf {\bibinfo {volume} {103}},\
  \bibinfo {pages} {122002} (\bibinfo {year} {2021}{\natexlab{a}})},\ \Eprint
  {http://arxiv.org/abs/2010.14529} {arXiv:2010.14529 [gr-qc]} \BibitemShut
  {NoStop}%
\bibitem [{\citenamefont {Abbott}\ \emph
  {et~al.}(2021{\natexlab{b}})\citenamefont {Abbott} \emph
  {et~al.}}]{LIGOScientific:2021djp}%
  \BibitemOpen
  \bibfield  {author} {\bibinfo {author} {\bibfnamefont {R.}~\bibnamefont
  {Abbott}} \emph {et~al.} (\bibinfo {collaboration} {LIGO Scientific, VIRGO,
  KAGRA}),\ }\href@noop {} {\  (\bibinfo {year} {2021}{\natexlab{b}})},\
  \Eprint {http://arxiv.org/abs/2111.03606} {arXiv:2111.03606 [gr-qc]}
  \BibitemShut {NoStop}%
\bibitem [{\citenamefont {Abbott}\ \emph
  {et~al.}(2021{\natexlab{c}})\citenamefont {Abbott} \emph
  {et~al.}}]{LIGOScientific:2021sio}%
  \BibitemOpen
  \bibfield  {author} {\bibinfo {author} {\bibfnamefont {R.}~\bibnamefont
  {Abbott}} \emph {et~al.} (\bibinfo {collaboration} {LIGO Scientific, VIRGO,
  KAGRA}),\ }\href@noop {} {\  (\bibinfo {year} {2021}{\natexlab{c}})},\
  \Eprint {http://arxiv.org/abs/2112.06861} {arXiv:2112.06861 [gr-qc]}
  \BibitemShut {NoStop}%
\bibitem [{\citenamefont {Akiyama}\ \emph
  {et~al.}(2019{\natexlab{a}})\citenamefont {Akiyama} \emph
  {et~al.}}]{EventHorizonTelescope:2019dse}%
  \BibitemOpen
  \bibfield  {author} {\bibinfo {author} {\bibfnamefont {K.}~\bibnamefont
  {Akiyama}} \emph {et~al.} (\bibinfo {collaboration} {Event Horizon
  Telescope}),\ }\href {\doibase 10.3847/2041-8213/ab0ec7} {\bibfield
  {journal} {\bibinfo  {journal} {Astrophys. J. Lett.}\ }\textbf {\bibinfo
  {volume} {875}},\ \bibinfo {pages} {L1} (\bibinfo {year}
  {2019}{\natexlab{a}})},\ \Eprint {http://arxiv.org/abs/1906.11238}
  {arXiv:1906.11238 [astro-ph.GA]} \BibitemShut {NoStop}%
\bibitem [{\citenamefont {Akiyama}\ \emph
  {et~al.}(2022{\natexlab{a}})\citenamefont {Akiyama} \emph
  {et~al.}}]{EventHorizonTelescope:2022wkp}%
  \BibitemOpen
  \bibfield  {author} {\bibinfo {author} {\bibfnamefont {K.}~\bibnamefont
  {Akiyama}} \emph {et~al.} (\bibinfo {collaboration} {Event Horizon
  Telescope}),\ }\href {\doibase 10.3847/2041-8213/ac6674} {\bibfield
  {journal} {\bibinfo  {journal} {Astrophys. J. Lett.}\ }\textbf {\bibinfo
  {volume} {930}},\ \bibinfo {pages} {L12} (\bibinfo {year}
  {2022}{\natexlab{a}})},\ \Eprint {http://arxiv.org/abs/2311.08680}
  {arXiv:2311.08680 [astro-ph.HE]} \BibitemShut {NoStop}%
\bibitem [{\citenamefont {Akiyama}\ \emph
  {et~al.}(2019{\natexlab{b}})\citenamefont {Akiyama} \emph
  {et~al.}}]{EventHorizonTelescope:2019ggy}%
  \BibitemOpen
  \bibfield  {author} {\bibinfo {author} {\bibfnamefont {K.}~\bibnamefont
  {Akiyama}} \emph {et~al.} (\bibinfo {collaboration} {Event Horizon
  Telescope}),\ }\href {\doibase 10.3847/2041-8213/ab1141} {\bibfield
  {journal} {\bibinfo  {journal} {Astrophys. J. Lett.}\ }\textbf {\bibinfo
  {volume} {875}},\ \bibinfo {pages} {L6} (\bibinfo {year}
  {2019}{\natexlab{b}})},\ \Eprint {http://arxiv.org/abs/1906.11243}
  {arXiv:1906.11243 [astro-ph.GA]} \BibitemShut {NoStop}%
\bibitem [{\citenamefont {Kerr}(1963)}]{Kerr:1963ud}%
  \BibitemOpen
  \bibfield  {author} {\bibinfo {author} {\bibfnamefont {R.~P.}\ \bibnamefont
  {Kerr}},\ }\href {\doibase 10.1103/PhysRevLett.11.237} {\bibfield  {journal}
  {\bibinfo  {journal} {Phys. Rev. Lett.}\ }\textbf {\bibinfo {volume} {11}},\
  \bibinfo {pages} {237} (\bibinfo {year} {1963})}\BibitemShut {NoStop}%
\bibitem [{\citenamefont {Bekenstein}(1996)}]{Bekenstein:1996pn}%
  \BibitemOpen
  \bibfield  {author} {\bibinfo {author} {\bibfnamefont {J.~D.}\ \bibnamefont
  {Bekenstein}},\ }in\ \href@noop {} {\emph {\bibinfo {booktitle} {{2nd
  International Sakharov Conference on Physics}}}}\ (\bibinfo {year} {1996})\
  pp.\ \bibinfo {pages} {216--219},\ \Eprint
  {http://arxiv.org/abs/gr-qc/9605059} {arXiv:gr-qc/9605059} \BibitemShut
  {NoStop}%
\bibitem [{\citenamefont {Chrusciel}\ \emph {et~al.}(2012)\citenamefont
  {Chrusciel}, \citenamefont {Lopes~Costa},\ and\ \citenamefont
  {Heusler}}]{Chrusciel:2012jk}%
  \BibitemOpen
  \bibfield  {author} {\bibinfo {author} {\bibfnamefont {P.~T.}\ \bibnamefont
  {Chrusciel}}, \bibinfo {author} {\bibfnamefont {J.}~\bibnamefont
  {Lopes~Costa}}, \ and\ \bibinfo {author} {\bibfnamefont {M.}~\bibnamefont
  {Heusler}},\ }\href {\doibase 10.12942/lrr-2012-7} {\bibfield  {journal}
  {\bibinfo  {journal} {Living Rev. Rel.}\ }\textbf {\bibinfo {volume} {15}},\
  \bibinfo {pages} {7} (\bibinfo {year} {2012})},\ \Eprint
  {http://arxiv.org/abs/1205.6112} {arXiv:1205.6112 [gr-qc]} \BibitemShut
  {NoStop}%
\bibitem [{\citenamefont {Gralla}\ \emph {et~al.}(2019)\citenamefont {Gralla},
  \citenamefont {Holz},\ and\ \citenamefont {Wald}}]{Gralla:2019xty}%
  \BibitemOpen
  \bibfield  {author} {\bibinfo {author} {\bibfnamefont {S.~E.}\ \bibnamefont
  {Gralla}}, \bibinfo {author} {\bibfnamefont {D.~E.}\ \bibnamefont {Holz}}, \
  and\ \bibinfo {author} {\bibfnamefont {R.~M.}\ \bibnamefont {Wald}},\ }\href
  {\doibase 10.1103/PhysRevD.100.024018} {\bibfield  {journal} {\bibinfo
  {journal} {Phys. Rev. D}\ }\textbf {\bibinfo {volume} {100}},\ \bibinfo
  {pages} {024018} (\bibinfo {year} {2019})},\ \Eprint
  {http://arxiv.org/abs/1906.00873} {arXiv:1906.00873 [astro-ph.HE]}
  \BibitemShut {NoStop}%
\bibitem [{\citenamefont {Akiyama}\ \emph
  {et~al.}(2022{\natexlab{b}})\citenamefont {Akiyama} \emph
  {et~al.}}]{EventHorizonTelescope:2022xqj}%
  \BibitemOpen
  \bibfield  {author} {\bibinfo {author} {\bibfnamefont {K.}~\bibnamefont
  {Akiyama}} \emph {et~al.} (\bibinfo {collaboration} {Event Horizon
  Telescope}),\ }\href {\doibase 10.3847/2041-8213/ac6756} {\bibfield
  {journal} {\bibinfo  {journal} {Astrophys. J. Lett.}\ }\textbf {\bibinfo
  {volume} {930}},\ \bibinfo {pages} {L17} (\bibinfo {year}
  {2022}{\natexlab{b}})},\ \Eprint {http://arxiv.org/abs/2311.09484}
  {arXiv:2311.09484 [astro-ph.HE]} \BibitemShut {NoStop}%
\bibitem [{\citenamefont {Abbott}\ \emph
  {et~al.}(2019{\natexlab{b}})\citenamefont {Abbott} \emph
  {et~al.}}]{LIGOScientific:2019fpa}%
  \BibitemOpen
  \bibfield  {author} {\bibinfo {author} {\bibfnamefont {B.~P.}\ \bibnamefont
  {Abbott}} \emph {et~al.} (\bibinfo {collaboration} {LIGO Scientific,
  Virgo}),\ }\href {\doibase 10.1103/PhysRevD.100.104036} {\bibfield  {journal}
  {\bibinfo  {journal} {Phys. Rev. D}\ }\textbf {\bibinfo {volume} {100}},\
  \bibinfo {pages} {104036} (\bibinfo {year} {2019}{\natexlab{b}})},\ \Eprint
  {http://arxiv.org/abs/1903.04467} {arXiv:1903.04467 [gr-qc]} \BibitemShut
  {NoStop}%
\bibitem [{\citenamefont {Ghosh}(2022)}]{Ghosh:2022xhn}%
  \BibitemOpen
  \bibfield  {author} {\bibinfo {author} {\bibfnamefont {A.}~\bibnamefont
  {Ghosh}} (\bibinfo {collaboration} {LIGO Scientific--Virgo--Kagra}),\ }in\
  \href@noop {} {\emph {\bibinfo {booktitle} {{56th Rencontres de Moriond on
  Gravitation}}}}\ (\bibinfo {year} {2022})\ \Eprint
  {http://arxiv.org/abs/2204.00662} {arXiv:2204.00662 [gr-qc]} \BibitemShut
  {NoStop}%
\bibitem [{\citenamefont {Psaltis}\ \emph {et~al.}(2020)\citenamefont {Psaltis}
  \emph {et~al.}}]{EventHorizonTelescope:2020qrl}%
  \BibitemOpen
  \bibfield  {author} {\bibinfo {author} {\bibfnamefont {D.}~\bibnamefont
  {Psaltis}} \emph {et~al.} (\bibinfo {collaboration} {Event Horizon
  Telescope}),\ }\href {\doibase 10.1103/PhysRevLett.125.141104} {\bibfield
  {journal} {\bibinfo  {journal} {Phys. Rev. Lett.}\ }\textbf {\bibinfo
  {volume} {125}},\ \bibinfo {pages} {141104} (\bibinfo {year} {2020})},\
  \Eprint {http://arxiv.org/abs/2010.01055} {arXiv:2010.01055 [gr-qc]}
  \BibitemShut {NoStop}%
\bibitem [{\citenamefont {Punturo}\ \emph {et~al.}(2010)\citenamefont {Punturo}
  \emph {et~al.}}]{Punturo:2010zz}%
  \BibitemOpen
  \bibfield  {author} {\bibinfo {author} {\bibfnamefont {M.}~\bibnamefont
  {Punturo}} \emph {et~al.},\ }\href {\doibase 10.1088/0264-9381/27/19/194002}
  {\bibfield  {journal} {\bibinfo  {journal} {Class. Quant. Grav.}\ }\textbf
  {\bibinfo {volume} {27}},\ \bibinfo {pages} {194002} (\bibinfo {year}
  {2010})}\BibitemShut {NoStop}%
\bibitem [{\citenamefont {Amaro-Seoane}\ \emph {et~al.}(2017)\citenamefont
  {Amaro-Seoane} \emph {et~al.}}]{LISA:2017pwj}%
  \BibitemOpen
  \bibfield  {author} {\bibinfo {author} {\bibfnamefont {P.}~\bibnamefont
  {Amaro-Seoane}} \emph {et~al.} (\bibinfo {collaboration} {LISA}),\
  }\href@noop {} {\  (\bibinfo {year} {2017})},\ \Eprint
  {http://arxiv.org/abs/1702.00786} {arXiv:1702.00786 [astro-ph.IM]}
  \BibitemShut {NoStop}%
\bibitem [{\citenamefont {Luo}\ \emph {et~al.}(2016)\citenamefont {Luo} \emph
  {et~al.}}]{TianQin:2015yph}%
  \BibitemOpen
  \bibfield  {author} {\bibinfo {author} {\bibfnamefont {J.}~\bibnamefont
  {Luo}} \emph {et~al.} (\bibinfo {collaboration} {TianQin}),\ }\href {\doibase
  10.1088/0264-9381/33/3/035010} {\bibfield  {journal} {\bibinfo  {journal}
  {Class. Quant. Grav.}\ }\textbf {\bibinfo {volume} {33}},\ \bibinfo {pages}
  {035010} (\bibinfo {year} {2016})},\ \Eprint
  {http://arxiv.org/abs/1512.02076} {arXiv:1512.02076 [astro-ph.IM]}
  \BibitemShut {NoStop}%
\bibitem [{\citenamefont {Gong}\ \emph {et~al.}(2021)\citenamefont {Gong},
  \citenamefont {Luo},\ and\ \citenamefont {Wang}}]{Gong:2021gvw}%
  \BibitemOpen
  \bibfield  {author} {\bibinfo {author} {\bibfnamefont {Y.}~\bibnamefont
  {Gong}}, \bibinfo {author} {\bibfnamefont {J.}~\bibnamefont {Luo}}, \ and\
  \bibinfo {author} {\bibfnamefont {B.}~\bibnamefont {Wang}},\ }\href {\doibase
  10.1038/s41550-021-01480-3} {\bibfield  {journal} {\bibinfo  {journal}
  {Nature Astron.}\ }\textbf {\bibinfo {volume} {5}},\ \bibinfo {pages} {881}
  (\bibinfo {year} {2021})},\ \Eprint {http://arxiv.org/abs/2109.07442}
  {arXiv:2109.07442 [astro-ph.IM]} \BibitemShut {NoStop}%
\bibitem [{\citenamefont {Hartle}(1973)}]{Hartle:1973zz}%
  \BibitemOpen
  \bibfield  {author} {\bibinfo {author} {\bibfnamefont {J.~B.}\ \bibnamefont
  {Hartle}},\ }\href {\doibase 10.1103/PhysRevD.8.1010} {\bibfield  {journal}
  {\bibinfo  {journal} {Phys. Rev. D}\ }\textbf {\bibinfo {volume} {8}},\
  \bibinfo {pages} {1010} (\bibinfo {year} {1973})}\BibitemShut {NoStop}%
\bibitem [{\citenamefont {Hughes}(2001)}]{Hughes:2001jr}%
  \BibitemOpen
  \bibfield  {author} {\bibinfo {author} {\bibfnamefont {S.~A.}\ \bibnamefont
  {Hughes}},\ }\href {\doibase 10.1103/PhysRevD.64.064004} {\bibfield
  {journal} {\bibinfo  {journal} {Phys. Rev. D}\ }\textbf {\bibinfo {volume}
  {64}},\ \bibinfo {pages} {064004} (\bibinfo {year} {2001})},\ \bibinfo {note}
  {[Erratum: Phys.Rev.D 88, 109902 (2013)]},\ \Eprint
  {http://arxiv.org/abs/gr-qc/0104041} {arXiv:gr-qc/0104041} \BibitemShut
  {NoStop}%
\bibitem [{\citenamefont {Comeau}\ and\ \citenamefont
  {Poisson}(2009)}]{Comeau:2009bz}%
  \BibitemOpen
  \bibfield  {author} {\bibinfo {author} {\bibfnamefont {S.}~\bibnamefont
  {Comeau}}\ and\ \bibinfo {author} {\bibfnamefont {E.}~\bibnamefont
  {Poisson}},\ }\href {\doibase 10.1103/PhysRevD.80.087501} {\bibfield
  {journal} {\bibinfo  {journal} {Phys. Rev. D}\ }\textbf {\bibinfo {volume}
  {80}},\ \bibinfo {pages} {087501} (\bibinfo {year} {2009})},\ \Eprint
  {http://arxiv.org/abs/0908.4518} {arXiv:0908.4518 [gr-qc]} \BibitemShut
  {NoStop}%
\bibitem [{\citenamefont {Poisson}(2015)}]{Poisson:2014gka}%
  \BibitemOpen
  \bibfield  {author} {\bibinfo {author} {\bibfnamefont {E.}~\bibnamefont
  {Poisson}},\ }\href {\doibase 10.1103/PhysRevD.91.044004} {\bibfield
  {journal} {\bibinfo  {journal} {Phys. Rev. D}\ }\textbf {\bibinfo {volume}
  {91}},\ \bibinfo {pages} {044004} (\bibinfo {year} {2015})},\ \Eprint
  {http://arxiv.org/abs/1411.4711} {arXiv:1411.4711 [gr-qc]} \BibitemShut
  {NoStop}%
\bibitem [{\citenamefont {Landry}\ and\ \citenamefont
  {Poisson}(2015)}]{Landry:2015zfa}%
  \BibitemOpen
  \bibfield  {author} {\bibinfo {author} {\bibfnamefont {P.}~\bibnamefont
  {Landry}}\ and\ \bibinfo {author} {\bibfnamefont {E.}~\bibnamefont
  {Poisson}},\ }\href {\doibase 10.1103/PhysRevD.91.104018} {\bibfield
  {journal} {\bibinfo  {journal} {Phys. Rev. D}\ }\textbf {\bibinfo {volume}
  {91}},\ \bibinfo {pages} {104018} (\bibinfo {year} {2015})},\ \Eprint
  {http://arxiv.org/abs/1503.07366} {arXiv:1503.07366 [gr-qc]} \BibitemShut
  {NoStop}%
\bibitem [{\citenamefont {Chatziioannou}\ \emph {et~al.}(2016)\citenamefont
  {Chatziioannou}, \citenamefont {Poisson},\ and\ \citenamefont
  {Yunes}}]{Chatziioannou:2016kem}%
  \BibitemOpen
  \bibfield  {author} {\bibinfo {author} {\bibfnamefont {K.}~\bibnamefont
  {Chatziioannou}}, \bibinfo {author} {\bibfnamefont {E.}~\bibnamefont
  {Poisson}}, \ and\ \bibinfo {author} {\bibfnamefont {N.}~\bibnamefont
  {Yunes}},\ }\href {\doibase 10.1103/PhysRevD.94.084043} {\bibfield  {journal}
  {\bibinfo  {journal} {Phys. Rev. D}\ }\textbf {\bibinfo {volume} {94}},\
  \bibinfo {pages} {084043} (\bibinfo {year} {2016})},\ \Eprint
  {http://arxiv.org/abs/1608.02899} {arXiv:1608.02899 [gr-qc]} \BibitemShut
  {NoStop}%
\bibitem [{\citenamefont {Pitre}\ and\ \citenamefont
  {Poisson}(2023)}]{Pitre:2023xsr}%
  \BibitemOpen
  \bibfield  {author} {\bibinfo {author} {\bibfnamefont {T.}~\bibnamefont
  {Pitre}}\ and\ \bibinfo {author} {\bibfnamefont {E.}~\bibnamefont
  {Poisson}},\ }\href@noop {} {\  (\bibinfo {year} {2023})},\ \Eprint
  {http://arxiv.org/abs/2311.04075} {arXiv:2311.04075 [gr-qc]} \BibitemShut
  {NoStop}%
\bibitem [{\citenamefont {Brito}\ \emph {et~al.}(2015)\citenamefont {Brito},
  \citenamefont {Cardoso},\ and\ \citenamefont {Pani}}]{Brito:2015oca}%
  \BibitemOpen
  \bibfield  {author} {\bibinfo {author} {\bibfnamefont {R.}~\bibnamefont
  {Brito}}, \bibinfo {author} {\bibfnamefont {V.}~\bibnamefont {Cardoso}}, \
  and\ \bibinfo {author} {\bibfnamefont {P.}~\bibnamefont {Pani}},\ }\href
  {\doibase 10.1007/978-3-319-19000-6} {\bibfield  {journal} {\bibinfo
  {journal} {Lect. Notes Phys.}\ }\textbf {\bibinfo {volume} {906}},\ \bibinfo
  {pages} {pp.1} (\bibinfo {year} {2015})},\ \Eprint
  {http://arxiv.org/abs/1501.06570} {arXiv:1501.06570 [gr-qc]} \BibitemShut
  {NoStop}%
\bibitem [{\citenamefont {Poisson}(2009)}]{Poisson:2009di}%
  \BibitemOpen
  \bibfield  {author} {\bibinfo {author} {\bibfnamefont {E.}~\bibnamefont
  {Poisson}},\ }\href {\doibase 10.1103/PhysRevD.80.064029} {\bibfield
  {journal} {\bibinfo  {journal} {Phys. Rev. D}\ }\textbf {\bibinfo {volume}
  {80}},\ \bibinfo {pages} {064029} (\bibinfo {year} {2009})},\ \Eprint
  {http://arxiv.org/abs/0907.0874} {arXiv:0907.0874 [gr-qc]} \BibitemShut
  {NoStop}%
\bibitem [{\citenamefont {Cardoso}\ and\ \citenamefont
  {Pani}(2013)}]{Cardoso:2012zn}%
  \BibitemOpen
  \bibfield  {author} {\bibinfo {author} {\bibfnamefont {V.}~\bibnamefont
  {Cardoso}}\ and\ \bibinfo {author} {\bibfnamefont {P.}~\bibnamefont {Pani}},\
  }\href {\doibase 10.1088/0264-9381/30/4/045011} {\bibfield  {journal}
  {\bibinfo  {journal} {Class. Quant. Grav.}\ }\textbf {\bibinfo {volume}
  {30}},\ \bibinfo {pages} {045011} (\bibinfo {year} {2013})},\ \Eprint
  {http://arxiv.org/abs/1205.3184} {arXiv:1205.3184 [gr-qc]} \BibitemShut
  {NoStop}%
\bibitem [{\citenamefont {O'Sullivan}\ and\ \citenamefont
  {Hughes}(2016)}]{OSullivan:2015lni}%
  \BibitemOpen
  \bibfield  {author} {\bibinfo {author} {\bibfnamefont {S.}~\bibnamefont
  {O'Sullivan}}\ and\ \bibinfo {author} {\bibfnamefont {S.~A.}\ \bibnamefont
  {Hughes}},\ }\href {\doibase 10.1103/PhysRevD.94.044057} {\bibfield
  {journal} {\bibinfo  {journal} {Phys. Rev. D}\ }\textbf {\bibinfo {volume}
  {94}},\ \bibinfo {pages} {044057} (\bibinfo {year} {2016})},\ \Eprint
  {http://arxiv.org/abs/1505.03809} {arXiv:1505.03809 [gr-qc]} \BibitemShut
  {NoStop}%
\bibitem [{\citenamefont {Datta}\ \emph {et~al.}(2020)\citenamefont {Datta},
  \citenamefont {Brito}, \citenamefont {Bose}, \citenamefont {Pani},\ and\
  \citenamefont {Hughes}}]{Datta:2019epe}%
  \BibitemOpen
  \bibfield  {author} {\bibinfo {author} {\bibfnamefont {S.}~\bibnamefont
  {Datta}}, \bibinfo {author} {\bibfnamefont {R.}~\bibnamefont {Brito}},
  \bibinfo {author} {\bibfnamefont {S.}~\bibnamefont {Bose}}, \bibinfo {author}
  {\bibfnamefont {P.}~\bibnamefont {Pani}}, \ and\ \bibinfo {author}
  {\bibfnamefont {S.~A.}\ \bibnamefont {Hughes}},\ }\href {\doibase
  10.1103/PhysRevD.101.044004} {\bibfield  {journal} {\bibinfo  {journal}
  {Phys. Rev. D}\ }\textbf {\bibinfo {volume} {101}},\ \bibinfo {pages}
  {044004} (\bibinfo {year} {2020})},\ \Eprint
  {http://arxiv.org/abs/1910.07841} {arXiv:1910.07841 [gr-qc]} \BibitemShut
  {NoStop}%
\bibitem [{\citenamefont {Johnson-Mcdaniel}\ \emph {et~al.}(2020)\citenamefont
  {Johnson-Mcdaniel}, \citenamefont {Mukherjee}, \citenamefont {Kashyap},
  \citenamefont {Ajith}, \citenamefont {Del~Pozzo},\ and\ \citenamefont
  {Vitale}}]{Johnson-Mcdaniel:2018cdu}%
  \BibitemOpen
  \bibfield  {author} {\bibinfo {author} {\bibfnamefont {N.~K.}\ \bibnamefont
  {Johnson-Mcdaniel}}, \bibinfo {author} {\bibfnamefont {A.}~\bibnamefont
  {Mukherjee}}, \bibinfo {author} {\bibfnamefont {R.}~\bibnamefont {Kashyap}},
  \bibinfo {author} {\bibfnamefont {P.}~\bibnamefont {Ajith}}, \bibinfo
  {author} {\bibfnamefont {W.}~\bibnamefont {Del~Pozzo}}, \ and\ \bibinfo
  {author} {\bibfnamefont {S.}~\bibnamefont {Vitale}},\ }\href {\doibase
  10.1103/PhysRevD.102.123010} {\bibfield  {journal} {\bibinfo  {journal}
  {Phys. Rev. D}\ }\textbf {\bibinfo {volume} {102}},\ \bibinfo {pages}
  {123010} (\bibinfo {year} {2020})},\ \Eprint
  {http://arxiv.org/abs/1804.08026} {arXiv:1804.08026 [gr-qc]} \BibitemShut
  {NoStop}%
\bibitem [{\citenamefont {Ryan}(1995)}]{Ryan:1995wh}%
  \BibitemOpen
  \bibfield  {author} {\bibinfo {author} {\bibfnamefont {F.~D.}\ \bibnamefont
  {Ryan}},\ }\href {\doibase 10.1103/PhysRevD.52.5707} {\bibfield  {journal}
  {\bibinfo  {journal} {Phys. Rev. D}\ }\textbf {\bibinfo {volume} {52}},\
  \bibinfo {pages} {5707} (\bibinfo {year} {1995})}\BibitemShut {NoStop}%
\bibitem [{\citenamefont {Berry}\ \emph {et~al.}(2019)\citenamefont {Berry},
  \citenamefont {Hughes}, \citenamefont {Sopuerta}, \citenamefont {Chua},
  \citenamefont {Heffernan}, \citenamefont {Holley-Bockelmann}, \citenamefont
  {Mihaylov}, \citenamefont {Miller},\ and\ \citenamefont
  {Sesana}}]{Berry:2019wgg}%
  \BibitemOpen
  \bibfield  {author} {\bibinfo {author} {\bibfnamefont {C.~P.~L.}\
  \bibnamefont {Berry}}, \bibinfo {author} {\bibfnamefont {S.~A.}\ \bibnamefont
  {Hughes}}, \bibinfo {author} {\bibfnamefont {C.~F.}\ \bibnamefont
  {Sopuerta}}, \bibinfo {author} {\bibfnamefont {A.~J.~K.}\ \bibnamefont
  {Chua}}, \bibinfo {author} {\bibfnamefont {A.}~\bibnamefont {Heffernan}},
  \bibinfo {author} {\bibfnamefont {K.}~\bibnamefont {Holley-Bockelmann}},
  \bibinfo {author} {\bibfnamefont {D.~P.}\ \bibnamefont {Mihaylov}}, \bibinfo
  {author} {\bibfnamefont {M.~C.}\ \bibnamefont {Miller}}, \ and\ \bibinfo
  {author} {\bibfnamefont {A.}~\bibnamefont {Sesana}},\ }\href@noop {} {\
  (\bibinfo {year} {2019})},\ \Eprint {http://arxiv.org/abs/1903.03686}
  {arXiv:1903.03686 [astro-ph.HE]} \BibitemShut {NoStop}%
\bibitem [{\citenamefont {Pani}\ and\ \citenamefont
  {Maselli}(2019)}]{Pani:2019cyc}%
  \BibitemOpen
  \bibfield  {author} {\bibinfo {author} {\bibfnamefont {P.}~\bibnamefont
  {Pani}}\ and\ \bibinfo {author} {\bibfnamefont {A.}~\bibnamefont {Maselli}},\
  }\href {\doibase 10.1142/S0218271819440012} {\bibfield  {journal} {\bibinfo
  {journal} {Int. J. Mod. Phys. D}\ }\textbf {\bibinfo {volume} {28}},\
  \bibinfo {pages} {1944001} (\bibinfo {year} {2019})},\ \Eprint
  {http://arxiv.org/abs/1905.03947} {arXiv:1905.03947 [gr-qc]} \BibitemShut
  {NoStop}%
\bibitem [{\citenamefont {De~Luca}\ \emph {et~al.}(2023)\citenamefont
  {De~Luca}, \citenamefont {Maselli},\ and\ \citenamefont
  {Pani}}]{DeLuca:2022xlz}%
  \BibitemOpen
  \bibfield  {author} {\bibinfo {author} {\bibfnamefont {V.}~\bibnamefont
  {De~Luca}}, \bibinfo {author} {\bibfnamefont {A.}~\bibnamefont {Maselli}}, \
  and\ \bibinfo {author} {\bibfnamefont {P.}~\bibnamefont {Pani}},\ }\href
  {\doibase 10.1103/PhysRevD.107.044058} {\bibfield  {journal} {\bibinfo
  {journal} {Phys. Rev. D}\ }\textbf {\bibinfo {volume} {107}},\ \bibinfo
  {pages} {044058} (\bibinfo {year} {2023})},\ \Eprint
  {http://arxiv.org/abs/2212.03343} {arXiv:2212.03343 [gr-qc]} \BibitemShut
  {NoStop}%
\bibitem [{\citenamefont {Zi}\ and\ \citenamefont {Li}(2023)}]{Zi:2023pvl}%
  \BibitemOpen
  \bibfield  {author} {\bibinfo {author} {\bibfnamefont {T.}~\bibnamefont
  {Zi}}\ and\ \bibinfo {author} {\bibfnamefont {P.-C.}\ \bibnamefont {Li}},\
  }\href {\doibase 10.1103/PhysRevD.108.024018} {\bibfield  {journal} {\bibinfo
   {journal} {Phys. Rev. D}\ }\textbf {\bibinfo {volume} {108}},\ \bibinfo
  {pages} {024018} (\bibinfo {year} {2023})},\ \Eprint
  {http://arxiv.org/abs/2303.16610} {arXiv:2303.16610 [gr-qc]} \BibitemShut
  {NoStop}%
\bibitem [{\citenamefont {Datta}\ and\ \citenamefont
  {Bose}(2019)}]{Datta:2019euh}%
  \BibitemOpen
  \bibfield  {author} {\bibinfo {author} {\bibfnamefont {S.}~\bibnamefont
  {Datta}}\ and\ \bibinfo {author} {\bibfnamefont {S.}~\bibnamefont {Bose}},\
  }\href {\doibase 10.1103/PhysRevD.99.084001} {\bibfield  {journal} {\bibinfo
  {journal} {Phys. Rev. D}\ }\textbf {\bibinfo {volume} {99}},\ \bibinfo
  {pages} {084001} (\bibinfo {year} {2019})},\ \Eprint
  {http://arxiv.org/abs/1902.01723} {arXiv:1902.01723 [gr-qc]} \BibitemShut
  {NoStop}%
\bibitem [{\citenamefont {Maggio}\ \emph {et~al.}(2021)\citenamefont {Maggio},
  \citenamefont {van~de Meent},\ and\ \citenamefont {Pani}}]{Maggio:2021uge}%
  \BibitemOpen
  \bibfield  {author} {\bibinfo {author} {\bibfnamefont {E.}~\bibnamefont
  {Maggio}}, \bibinfo {author} {\bibfnamefont {M.}~\bibnamefont {van~de
  Meent}}, \ and\ \bibinfo {author} {\bibfnamefont {P.}~\bibnamefont {Pani}},\
  }\href {\doibase 10.1103/PhysRevD.104.104026} {\bibfield  {journal} {\bibinfo
   {journal} {Phys. Rev. D}\ }\textbf {\bibinfo {volume} {104}},\ \bibinfo
  {pages} {104026} (\bibinfo {year} {2021})},\ \Eprint
  {http://arxiv.org/abs/2106.07195} {arXiv:2106.07195 [gr-qc]} \BibitemShut
  {NoStop}%
\bibitem [{\citenamefont {Sago}\ and\ \citenamefont
  {Tanaka}(2021)}]{Sago:2021iku}%
  \BibitemOpen
  \bibfield  {author} {\bibinfo {author} {\bibfnamefont {N.}~\bibnamefont
  {Sago}}\ and\ \bibinfo {author} {\bibfnamefont {T.}~\bibnamefont {Tanaka}},\
  }\href {\doibase 10.1103/PhysRevD.104.064009} {\bibfield  {journal} {\bibinfo
   {journal} {Phys. Rev. D}\ }\textbf {\bibinfo {volume} {104}},\ \bibinfo
  {pages} {064009} (\bibinfo {year} {2021})},\ \Eprint
  {http://arxiv.org/abs/2106.07123} {arXiv:2106.07123 [gr-qc]} \BibitemShut
  {NoStop}%
\bibitem [{\citenamefont {Sago}\ and\ \citenamefont
  {Tanaka}(2022)}]{Sago:2022bbj}%
  \BibitemOpen
  \bibfield  {author} {\bibinfo {author} {\bibfnamefont {N.}~\bibnamefont
  {Sago}}\ and\ \bibinfo {author} {\bibfnamefont {T.}~\bibnamefont {Tanaka}},\
  }\href {\doibase 10.1103/PhysRevD.106.024032} {\bibfield  {journal} {\bibinfo
   {journal} {Phys. Rev. D}\ }\textbf {\bibinfo {volume} {106}},\ \bibinfo
  {pages} {024032} (\bibinfo {year} {2022})},\ \Eprint
  {http://arxiv.org/abs/2202.04249} {arXiv:2202.04249 [gr-qc]} \BibitemShut
  {NoStop}%
\bibitem [{\citenamefont {Cutler}\ and\ \citenamefont
  {Flanagan}(1994)}]{Cutler:1994ys}%
  \BibitemOpen
  \bibfield  {author} {\bibinfo {author} {\bibfnamefont {C.}~\bibnamefont
  {Cutler}}\ and\ \bibinfo {author} {\bibfnamefont {E.~E.}\ \bibnamefont
  {Flanagan}},\ }\href {\doibase 10.1103/PhysRevD.49.2658} {\bibfield
  {journal} {\bibinfo  {journal} {Phys. Rev. D}\ }\textbf {\bibinfo {volume}
  {49}},\ \bibinfo {pages} {2658} (\bibinfo {year} {1994})},\ \Eprint
  {http://arxiv.org/abs/gr-qc/9402014} {arXiv:gr-qc/9402014} \BibitemShut
  {NoStop}%
\bibitem [{\citenamefont {Vallisneri}(2008)}]{Vallisneri:2007ev}%
  \BibitemOpen
  \bibfield  {author} {\bibinfo {author} {\bibfnamefont {M.}~\bibnamefont
  {Vallisneri}},\ }\href {\doibase 10.1103/PhysRevD.77.042001} {\bibfield
  {journal} {\bibinfo  {journal} {Phys. Rev. D}\ }\textbf {\bibinfo {volume}
  {77}},\ \bibinfo {pages} {042001} (\bibinfo {year} {2008})},\ \Eprint
  {http://arxiv.org/abs/gr-qc/0703086} {arXiv:gr-qc/0703086} \BibitemShut
  {NoStop}%
\bibitem [{\citenamefont {Sago}\ and\ \citenamefont
  {Fujita}(2015)}]{Sago:2015rpa}%
  \BibitemOpen
  \bibfield  {author} {\bibinfo {author} {\bibfnamefont {N.}~\bibnamefont
  {Sago}}\ and\ \bibinfo {author} {\bibfnamefont {R.}~\bibnamefont {Fujita}},\
  }\href {\doibase 10.1093/ptep/ptv092} {\bibfield  {journal} {\bibinfo
  {journal} {PTEP}\ }\textbf {\bibinfo {volume} {2015}},\ \bibinfo {pages}
  {073E03} (\bibinfo {year} {2015})},\ \Eprint
  {http://arxiv.org/abs/1505.01600} {arXiv:1505.01600 [gr-qc]} \BibitemShut
  {NoStop}%
\bibitem [{\citenamefont {Munna}\ \emph {et~al.}(2020)\citenamefont {Munna},
  \citenamefont {Evans}, \citenamefont {Hopper},\ and\ \citenamefont
  {Forseth}}]{Munna:2020juq}%
  \BibitemOpen
  \bibfield  {author} {\bibinfo {author} {\bibfnamefont {C.}~\bibnamefont
  {Munna}}, \bibinfo {author} {\bibfnamefont {C.~R.}\ \bibnamefont {Evans}},
  \bibinfo {author} {\bibfnamefont {S.}~\bibnamefont {Hopper}}, \ and\ \bibinfo
  {author} {\bibfnamefont {E.}~\bibnamefont {Forseth}},\ }\href {\doibase
  10.1103/PhysRevD.102.024047} {\bibfield  {journal} {\bibinfo  {journal}
  {Phys. Rev. D}\ }\textbf {\bibinfo {volume} {102}},\ \bibinfo {pages}
  {024047} (\bibinfo {year} {2020})},\ \Eprint
  {http://arxiv.org/abs/2005.03044} {arXiv:2005.03044 [gr-qc]} \BibitemShut
  {NoStop}%
\bibitem [{\citenamefont {Munna}\ \emph {et~al.}(2023)\citenamefont {Munna},
  \citenamefont {Evans},\ and\ \citenamefont {Forseth}}]{Munna:2023vds}%
  \BibitemOpen
  \bibfield  {author} {\bibinfo {author} {\bibfnamefont {C.}~\bibnamefont
  {Munna}}, \bibinfo {author} {\bibfnamefont {C.~R.}\ \bibnamefont {Evans}}, \
  and\ \bibinfo {author} {\bibfnamefont {E.}~\bibnamefont {Forseth}},\ }\href
  {\doibase 10.1103/PhysRevD.108.044039} {\bibfield  {journal} {\bibinfo
  {journal} {Phys. Rev. D}\ }\textbf {\bibinfo {volume} {108}},\ \bibinfo
  {pages} {044039} (\bibinfo {year} {2023})},\ \Eprint
  {http://arxiv.org/abs/2306.12481} {arXiv:2306.12481 [gr-qc]} \BibitemShut
  {NoStop}%
\bibitem [{\citenamefont {Katz}\ \emph {et~al.}(2021)\citenamefont {Katz},
  \citenamefont {Chua}, \citenamefont {Speri}, \citenamefont {Warburton},\ and\
  \citenamefont {Hughes}}]{Katz:2021yft}%
  \BibitemOpen
  \bibfield  {author} {\bibinfo {author} {\bibfnamefont {M.~L.}\ \bibnamefont
  {Katz}}, \bibinfo {author} {\bibfnamefont {A.~J.~K.}\ \bibnamefont {Chua}},
  \bibinfo {author} {\bibfnamefont {L.}~\bibnamefont {Speri}}, \bibinfo
  {author} {\bibfnamefont {N.}~\bibnamefont {Warburton}}, \ and\ \bibinfo
  {author} {\bibfnamefont {S.~A.}\ \bibnamefont {Hughes}},\ }\href@noop {} {\
  (\bibinfo {year} {2021})},\ \Eprint {http://arxiv.org/abs/2104.04582}
  {arXiv:2104.04582 [gr-qc]} \BibitemShut {NoStop}%
\bibitem [{\citenamefont {Maggio}\ \emph {et~al.}(2017)\citenamefont {Maggio},
  \citenamefont {Pani},\ and\ \citenamefont {Ferrari}}]{Maggio:2017ivp}%
  \BibitemOpen
  \bibfield  {author} {\bibinfo {author} {\bibfnamefont {E.}~\bibnamefont
  {Maggio}}, \bibinfo {author} {\bibfnamefont {P.}~\bibnamefont {Pani}}, \ and\
  \bibinfo {author} {\bibfnamefont {V.}~\bibnamefont {Ferrari}},\ }\href
  {\doibase 10.1103/PhysRevD.96.104047} {\bibfield  {journal} {\bibinfo
  {journal} {Phys. Rev. D}\ }\textbf {\bibinfo {volume} {96}},\ \bibinfo
  {pages} {104047} (\bibinfo {year} {2017})},\ \Eprint
  {http://arxiv.org/abs/1703.03696} {arXiv:1703.03696 [gr-qc]} \BibitemShut
  {NoStop}%
\bibitem [{\citenamefont {Maggio}\ \emph {et~al.}(2019)\citenamefont {Maggio},
  \citenamefont {Cardoso}, \citenamefont {Dolan},\ and\ \citenamefont
  {Pani}}]{Maggio:2018ivz}%
  \BibitemOpen
  \bibfield  {author} {\bibinfo {author} {\bibfnamefont {E.}~\bibnamefont
  {Maggio}}, \bibinfo {author} {\bibfnamefont {V.}~\bibnamefont {Cardoso}},
  \bibinfo {author} {\bibfnamefont {S.~R.}\ \bibnamefont {Dolan}}, \ and\
  \bibinfo {author} {\bibfnamefont {P.}~\bibnamefont {Pani}},\ }\href {\doibase
  10.1103/PhysRevD.99.064007} {\bibfield  {journal} {\bibinfo  {journal} {Phys.
  Rev. D}\ }\textbf {\bibinfo {volume} {99}},\ \bibinfo {pages} {064007}
  (\bibinfo {year} {2019})},\ \Eprint {http://arxiv.org/abs/1807.08840}
  {arXiv:1807.08840 [gr-qc]} \BibitemShut {NoStop}%
\bibitem [{\citenamefont {Wang}\ and\ \citenamefont
  {Afshordi}(2018)}]{Wang:2018gin}%
  \BibitemOpen
  \bibfield  {author} {\bibinfo {author} {\bibfnamefont {Q.}~\bibnamefont
  {Wang}}\ and\ \bibinfo {author} {\bibfnamefont {N.}~\bibnamefont
  {Afshordi}},\ }\href {\doibase 10.1103/PhysRevD.97.124044} {\bibfield
  {journal} {\bibinfo  {journal} {Phys. Rev. D}\ }\textbf {\bibinfo {volume}
  {97}},\ \bibinfo {pages} {124044} (\bibinfo {year} {2018})},\ \Eprint
  {http://arxiv.org/abs/1803.02845} {arXiv:1803.02845 [gr-qc]} \BibitemShut
  {NoStop}%
\bibitem [{\citenamefont {Yagi}\ and\ \citenamefont
  {Yunes}(2015)}]{Yagi:2015upa}%
  \BibitemOpen
  \bibfield  {author} {\bibinfo {author} {\bibfnamefont {K.}~\bibnamefont
  {Yagi}}\ and\ \bibinfo {author} {\bibfnamefont {N.}~\bibnamefont {Yunes}},\
  }\href {\doibase 10.1103/PhysRevD.91.103003} {\bibfield  {journal} {\bibinfo
  {journal} {Phys. Rev. D}\ }\textbf {\bibinfo {volume} {91}},\ \bibinfo
  {pages} {103003} (\bibinfo {year} {2015})},\ \Eprint
  {http://arxiv.org/abs/1502.04131} {arXiv:1502.04131 [gr-qc]} \BibitemShut
  {NoStop}%
\bibitem [{\citenamefont {Pani}(2015)}]{Pani:2015tga}%
  \BibitemOpen
  \bibfield  {author} {\bibinfo {author} {\bibfnamefont {P.}~\bibnamefont
  {Pani}},\ }\href {\doibase 10.1103/PhysRevD.95.049902} {\bibfield  {journal}
  {\bibinfo  {journal} {Phys. Rev. D}\ }\textbf {\bibinfo {volume} {92}},\
  \bibinfo {pages} {124030} (\bibinfo {year} {2015})},\ \bibinfo {note}
  {[Erratum: Phys.Rev.D 95, 049902 (2017)]},\ \Eprint
  {http://arxiv.org/abs/1506.06050} {arXiv:1506.06050 [gr-qc]} \BibitemShut
  {NoStop}%
\bibitem [{\citenamefont {Raposo}\ \emph {et~al.}(2019)\citenamefont {Raposo},
  \citenamefont {Pani},\ and\ \citenamefont {Emparan}}]{Raposo:2018xkf}%
  \BibitemOpen
  \bibfield  {author} {\bibinfo {author} {\bibfnamefont {G.}~\bibnamefont
  {Raposo}}, \bibinfo {author} {\bibfnamefont {P.}~\bibnamefont {Pani}}, \ and\
  \bibinfo {author} {\bibfnamefont {R.}~\bibnamefont {Emparan}},\ }\href
  {\doibase 10.1103/PhysRevD.99.104050} {\bibfield  {journal} {\bibinfo
  {journal} {Phys. Rev. D}\ }\textbf {\bibinfo {volume} {99}},\ \bibinfo
  {pages} {104050} (\bibinfo {year} {2019})},\ \Eprint
  {http://arxiv.org/abs/1812.07615} {arXiv:1812.07615 [gr-qc]} \BibitemShut
  {NoStop}%
\bibitem [{\citenamefont {Thorne}\ \emph {et~al.}(1973)\citenamefont {Thorne},
  \citenamefont {Lee},\ and\ \citenamefont {Lightman}}]{Thorne:1973zz}%
  \BibitemOpen
  \bibfield  {author} {\bibinfo {author} {\bibfnamefont {K.~S.}\ \bibnamefont
  {Thorne}}, \bibinfo {author} {\bibfnamefont {D.~L.}\ \bibnamefont {Lee}}, \
  and\ \bibinfo {author} {\bibfnamefont {A.~P.}\ \bibnamefont {Lightman}},\
  }\href {\doibase 10.1103/PhysRevD.7.3563} {\bibfield  {journal} {\bibinfo
  {journal} {Phys. Rev. D}\ }\textbf {\bibinfo {volume} {7}},\ \bibinfo {pages}
  {3563} (\bibinfo {year} {1973})}\BibitemShut {NoStop}%
\bibitem [{\citenamefont {Drasco}\ and\ \citenamefont
  {Hughes}(2006)}]{Drasco:2005kz}%
  \BibitemOpen
  \bibfield  {author} {\bibinfo {author} {\bibfnamefont {S.}~\bibnamefont
  {Drasco}}\ and\ \bibinfo {author} {\bibfnamefont {S.~A.}\ \bibnamefont
  {Hughes}},\ }\href {\doibase 10.1103/PhysRevD.73.024027} {\bibfield
  {journal} {\bibinfo  {journal} {Phys. Rev. D}\ }\textbf {\bibinfo {volume}
  {73}},\ \bibinfo {pages} {024027} (\bibinfo {year} {2006})},\ \bibinfo {note}
  {[Erratum: Phys.Rev.D 88, 109905 (2013), Erratum: Phys.Rev.D 90, 109905
  (2014)]},\ \Eprint {http://arxiv.org/abs/gr-qc/0509101} {arXiv:gr-qc/0509101}
  \BibitemShut {NoStop}%
\bibitem [{\citenamefont {Schmidt}(2002)}]{Schmidt:2002qk}%
  \BibitemOpen
  \bibfield  {author} {\bibinfo {author} {\bibfnamefont {W.}~\bibnamefont
  {Schmidt}},\ }\href {\doibase 10.1088/0264-9381/19/10/314} {\bibfield
  {journal} {\bibinfo  {journal} {Class. Quant. Grav.}\ }\textbf {\bibinfo
  {volume} {19}},\ \bibinfo {pages} {2743} (\bibinfo {year} {2002})},\ \Eprint
  {http://arxiv.org/abs/gr-qc/0202090} {arXiv:gr-qc/0202090} \BibitemShut
  {NoStop}%
\bibitem [{\citenamefont {Fujita}\ and\ \citenamefont
  {Hikida}(2009)}]{Fujita:2009bp}%
  \BibitemOpen
  \bibfield  {author} {\bibinfo {author} {\bibfnamefont {R.}~\bibnamefont
  {Fujita}}\ and\ \bibinfo {author} {\bibfnamefont {W.}~\bibnamefont
  {Hikida}},\ }\href {\doibase 10.1088/0264-9381/26/13/135002} {\bibfield
  {journal} {\bibinfo  {journal} {Class. Quant. Grav.}\ }\textbf {\bibinfo
  {volume} {26}},\ \bibinfo {pages} {135002} (\bibinfo {year} {2009})},\
  \Eprint {http://arxiv.org/abs/0906.1420} {arXiv:0906.1420 [gr-qc]}
  \BibitemShut {NoStop}%
\bibitem [{\citenamefont {Hughes}\ \emph {et~al.}(2021)\citenamefont {Hughes},
  \citenamefont {Warburton}, \citenamefont {Khanna}, \citenamefont {Chua},\
  and\ \citenamefont {Katz}}]{Hughes:2021exa}%
  \BibitemOpen
  \bibfield  {author} {\bibinfo {author} {\bibfnamefont {S.~A.}\ \bibnamefont
  {Hughes}}, \bibinfo {author} {\bibfnamefont {N.}~\bibnamefont {Warburton}},
  \bibinfo {author} {\bibfnamefont {G.}~\bibnamefont {Khanna}}, \bibinfo
  {author} {\bibfnamefont {A.~J.~K.}\ \bibnamefont {Chua}}, \ and\ \bibinfo
  {author} {\bibfnamefont {M.~L.}\ \bibnamefont {Katz}},\ }\href {\doibase
  10.1103/PhysRevD.103.104014} {\bibfield  {journal} {\bibinfo  {journal}
  {Phys. Rev. D}\ }\textbf {\bibinfo {volume} {103}},\ \bibinfo {pages}
  {104014} (\bibinfo {year} {2021})},\ \bibinfo {note} {[Erratum: Phys.Rev.D
  107, 089901 (2023)]},\ \Eprint {http://arxiv.org/abs/2102.02713}
  {arXiv:2102.02713 [gr-qc]} \BibitemShut {NoStop}%
\bibitem [{\citenamefont {Isoyama}\ \emph {et~al.}(2022)\citenamefont
  {Isoyama}, \citenamefont {Fujita}, \citenamefont {Chua}, \citenamefont
  {Nakano}, \citenamefont {Pound},\ and\ \citenamefont
  {Sago}}]{Isoyama:2021jjd}%
  \BibitemOpen
  \bibfield  {author} {\bibinfo {author} {\bibfnamefont {S.}~\bibnamefont
  {Isoyama}}, \bibinfo {author} {\bibfnamefont {R.}~\bibnamefont {Fujita}},
  \bibinfo {author} {\bibfnamefont {A.~J.~K.}\ \bibnamefont {Chua}}, \bibinfo
  {author} {\bibfnamefont {H.}~\bibnamefont {Nakano}}, \bibinfo {author}
  {\bibfnamefont {A.}~\bibnamefont {Pound}}, \ and\ \bibinfo {author}
  {\bibfnamefont {N.}~\bibnamefont {Sago}},\ }\href {\doibase
  10.1103/PhysRevLett.128.231101} {\bibfield  {journal} {\bibinfo  {journal}
  {Phys. Rev. Lett.}\ }\textbf {\bibinfo {volume} {128}},\ \bibinfo {pages}
  {231101} (\bibinfo {year} {2022})},\ \Eprint
  {http://arxiv.org/abs/2111.05288} {arXiv:2111.05288 [gr-qc]} \BibitemShut
  {NoStop}%
\bibitem [{\citenamefont {Mark}\ \emph {et~al.}(2017)\citenamefont {Mark},
  \citenamefont {Zimmerman}, \citenamefont {Du},\ and\ \citenamefont
  {Chen}}]{Mark:2017dnq}%
  \BibitemOpen
  \bibfield  {author} {\bibinfo {author} {\bibfnamefont {Z.}~\bibnamefont
  {Mark}}, \bibinfo {author} {\bibfnamefont {A.}~\bibnamefont {Zimmerman}},
  \bibinfo {author} {\bibfnamefont {S.~M.}\ \bibnamefont {Du}}, \ and\ \bibinfo
  {author} {\bibfnamefont {Y.}~\bibnamefont {Chen}},\ }\href {\doibase
  10.1103/PhysRevD.96.084002} {\bibfield  {journal} {\bibinfo  {journal} {Phys.
  Rev. D}\ }\textbf {\bibinfo {volume} {96}},\ \bibinfo {pages} {084002}
  (\bibinfo {year} {2017})},\ \Eprint {http://arxiv.org/abs/1706.06155}
  {arXiv:1706.06155 [gr-qc]} \BibitemShut {NoStop}%
\bibitem [{\citenamefont {van~de Meent}(2020)}]{vandeMeent:2019cam}%
  \BibitemOpen
  \bibfield  {author} {\bibinfo {author} {\bibfnamefont {M.}~\bibnamefont
  {van~de Meent}},\ }\href {\doibase 10.1088/1361-6382/ab79d5} {\bibfield
  {journal} {\bibinfo  {journal} {Class. Quant. Grav.}\ }\textbf {\bibinfo
  {volume} {37}},\ \bibinfo {pages} {145007} (\bibinfo {year} {2020})},\
  \Eprint {http://arxiv.org/abs/1906.05090} {arXiv:1906.05090 [gr-qc]}
  \BibitemShut {NoStop}%
\bibitem [{\citenamefont {Stein}\ and\ \citenamefont
  {Warburton}(2020)}]{Stein:2019buj}%
  \BibitemOpen
  \bibfield  {author} {\bibinfo {author} {\bibfnamefont {L.~C.}\ \bibnamefont
  {Stein}}\ and\ \bibinfo {author} {\bibfnamefont {N.}~\bibnamefont
  {Warburton}},\ }\href {\doibase 10.1103/PhysRevD.101.064007} {\bibfield
  {journal} {\bibinfo  {journal} {Phys. Rev. D}\ }\textbf {\bibinfo {volume}
  {101}},\ \bibinfo {pages} {064007} (\bibinfo {year} {2020})},\ \Eprint
  {http://arxiv.org/abs/1912.07609} {arXiv:1912.07609 [gr-qc]} \BibitemShut
  {NoStop}%
\bibitem [{\citenamefont {Katz}\ \emph {et~al.}(2023)\citenamefont {Katz},
  \citenamefont {Speri}, \citenamefont {Chua}, \citenamefont {Chapman-Bird},
  \citenamefont {Warburton},\ and\ \citenamefont
  {Hughes}}]{michael_l_katz_2023_8190418}%
  \BibitemOpen
  \bibfield  {author} {\bibinfo {author} {\bibfnamefont {M.~L.}\ \bibnamefont
  {Katz}}, \bibinfo {author} {\bibfnamefont {L.}~\bibnamefont {Speri}},
  \bibinfo {author} {\bibfnamefont {A.~J.~K.}\ \bibnamefont {Chua}}, \bibinfo
  {author} {\bibfnamefont {C.~E.~A.}\ \bibnamefont {Chapman-Bird}}, \bibinfo
  {author} {\bibfnamefont {N.}~\bibnamefont {Warburton}}, \ and\ \bibinfo
  {author} {\bibfnamefont {S.~A.}\ \bibnamefont {Hughes}},\ }\href {\doibase
  10.5281/zenodo.8190418} {\enquote {\bibinfo {title}
  {{BlackHolePerturbationToolkit/FastEMRIWaveforms: Frequency Domain Waveform
  Added!}}}\ } (\bibinfo {year} {2023})\BibitemShut {NoStop}%
\bibitem [{\citenamefont {Chua}\ \emph {et~al.}(2017)\citenamefont {Chua},
  \citenamefont {Moore},\ and\ \citenamefont {Gair}}]{Chua:2017ujo}%
  \BibitemOpen
  \bibfield  {author} {\bibinfo {author} {\bibfnamefont {A.~J.}\ \bibnamefont
  {Chua}}, \bibinfo {author} {\bibfnamefont {C.~J.}\ \bibnamefont {Moore}}, \
  and\ \bibinfo {author} {\bibfnamefont {J.~R.}\ \bibnamefont {Gair}},\ }\href
  {\doibase 10.1103/PhysRevD.96.044005} {\bibfield  {journal} {\bibinfo
  {journal} {Phys. Rev. D}\ }\textbf {\bibinfo {volume} {96}},\ \bibinfo
  {pages} {044005} (\bibinfo {year} {2017})},\ \Eprint
  {http://arxiv.org/abs/1705.04259} {arXiv:1705.04259 [gr-qc]} \BibitemShut
  {NoStop}%
\bibitem [{\citenamefont {Speri}\ \emph {et~al.}(2023)\citenamefont {Speri},
  \citenamefont {Katz}, \citenamefont {Chua}, \citenamefont {Hughes},
  \citenamefont {Warburton}, \citenamefont {Thompson}, \citenamefont
  {Chapman-Bird},\ and\ \citenamefont {Gair}}]{Speri:2023jte}%
  \BibitemOpen
  \bibfield  {author} {\bibinfo {author} {\bibfnamefont {L.}~\bibnamefont
  {Speri}}, \bibinfo {author} {\bibfnamefont {M.~L.}\ \bibnamefont {Katz}},
  \bibinfo {author} {\bibfnamefont {A.~J.~K.}\ \bibnamefont {Chua}}, \bibinfo
  {author} {\bibfnamefont {S.~A.}\ \bibnamefont {Hughes}}, \bibinfo {author}
  {\bibfnamefont {N.}~\bibnamefont {Warburton}}, \bibinfo {author}
  {\bibfnamefont {J.~E.}\ \bibnamefont {Thompson}}, \bibinfo {author}
  {\bibfnamefont {C.~E.~A.}\ \bibnamefont {Chapman-Bird}}, \ and\ \bibinfo
  {author} {\bibfnamefont {J.~R.}\ \bibnamefont {Gair}},\ }\href@noop {} {\
  (\bibinfo {year} {2023})},\ \Eprint {http://arxiv.org/abs/2307.12585}
  {arXiv:2307.12585 [gr-qc]} \BibitemShut {NoStop}%
\bibitem [{\citenamefont {Speeney}\ \emph {et~al.}(2022)\citenamefont
  {Speeney}, \citenamefont {Antonelli}, \citenamefont {Baibhav},\ and\
  \citenamefont {Berti}}]{Speeney:2022ryg}%
  \BibitemOpen
  \bibfield  {author} {\bibinfo {author} {\bibfnamefont {N.}~\bibnamefont
  {Speeney}}, \bibinfo {author} {\bibfnamefont {A.}~\bibnamefont {Antonelli}},
  \bibinfo {author} {\bibfnamefont {V.}~\bibnamefont {Baibhav}}, \ and\
  \bibinfo {author} {\bibfnamefont {E.}~\bibnamefont {Berti}},\ }\href
  {\doibase 10.1103/PhysRevD.106.044027} {\bibfield  {journal} {\bibinfo
  {journal} {Phys. Rev. D}\ }\textbf {\bibinfo {volume} {106}},\ \bibinfo
  {pages} {044027} (\bibinfo {year} {2022})},\ \Eprint
  {http://arxiv.org/abs/2204.12508} {arXiv:2204.12508 [gr-qc]} \BibitemShut
  {NoStop}%
\bibitem [{\citenamefont {Flanagan}\ and\ \citenamefont
  {Hughes}(1998)}]{Flanagan:1997kp}%
  \BibitemOpen
  \bibfield  {author} {\bibinfo {author} {\bibfnamefont {E.~E.}\ \bibnamefont
  {Flanagan}}\ and\ \bibinfo {author} {\bibfnamefont {S.~A.}\ \bibnamefont
  {Hughes}},\ }\href {\doibase 10.1103/PhysRevD.57.4566} {\bibfield  {journal}
  {\bibinfo  {journal} {Phys. Rev. D}\ }\textbf {\bibinfo {volume} {57}},\
  \bibinfo {pages} {4566} (\bibinfo {year} {1998})},\ \Eprint
  {http://arxiv.org/abs/gr-qc/9710129} {arXiv:gr-qc/9710129} \BibitemShut
  {NoStop}%
\bibitem [{\citenamefont {Lindblom}\ \emph {et~al.}(2008)\citenamefont
  {Lindblom}, \citenamefont {Owen},\ and\ \citenamefont
  {Brown}}]{Lindblom:2008cm}%
  \BibitemOpen
  \bibfield  {author} {\bibinfo {author} {\bibfnamefont {L.}~\bibnamefont
  {Lindblom}}, \bibinfo {author} {\bibfnamefont {B.~J.}\ \bibnamefont {Owen}},
  \ and\ \bibinfo {author} {\bibfnamefont {D.~A.}\ \bibnamefont {Brown}},\
  }\href {\doibase 10.1103/PhysRevD.78.124020} {\bibfield  {journal} {\bibinfo
  {journal} {Phys. Rev. D}\ }\textbf {\bibinfo {volume} {78}},\ \bibinfo
  {pages} {124020} (\bibinfo {year} {2008})},\ \Eprint
  {http://arxiv.org/abs/0809.3844} {arXiv:0809.3844 [gr-qc]} \BibitemShut
  {NoStop}%
\bibitem [{\citenamefont {Babak}\ \emph {et~al.}(2017)\citenamefont {Babak},
  \citenamefont {Gair}, \citenamefont {Sesana}, \citenamefont {Barausse},
  \citenamefont {Sopuerta}, \citenamefont {Berry}, \citenamefont {Berti},
  \citenamefont {Amaro-Seoane}, \citenamefont {Petiteau},\ and\ \citenamefont
  {Klein}}]{Babak:2017tow}%
  \BibitemOpen
  \bibfield  {author} {\bibinfo {author} {\bibfnamefont {S.}~\bibnamefont
  {Babak}}, \bibinfo {author} {\bibfnamefont {J.}~\bibnamefont {Gair}},
  \bibinfo {author} {\bibfnamefont {A.}~\bibnamefont {Sesana}}, \bibinfo
  {author} {\bibfnamefont {E.}~\bibnamefont {Barausse}}, \bibinfo {author}
  {\bibfnamefont {C.~F.}\ \bibnamefont {Sopuerta}}, \bibinfo {author}
  {\bibfnamefont {C.~P.~L.}\ \bibnamefont {Berry}}, \bibinfo {author}
  {\bibfnamefont {E.}~\bibnamefont {Berti}}, \bibinfo {author} {\bibfnamefont
  {P.}~\bibnamefont {Amaro-Seoane}}, \bibinfo {author} {\bibfnamefont
  {A.}~\bibnamefont {Petiteau}}, \ and\ \bibinfo {author} {\bibfnamefont
  {A.}~\bibnamefont {Klein}},\ }\href {\doibase 10.1103/PhysRevD.95.103012}
  {\bibfield  {journal} {\bibinfo  {journal} {Phys. Rev. D}\ }\textbf {\bibinfo
  {volume} {95}},\ \bibinfo {pages} {103012} (\bibinfo {year} {2017})},\
  \Eprint {http://arxiv.org/abs/1703.09722} {arXiv:1703.09722 [gr-qc]}
  \BibitemShut {NoStop}%
\bibitem [{\citenamefont {Fan}\ \emph {et~al.}(2020)\citenamefont {Fan},
  \citenamefont {Hu}, \citenamefont {Barausse}, \citenamefont {Sesana},
  \citenamefont {Zhang}, \citenamefont {Zhang}, \citenamefont {Zi},\ and\
  \citenamefont {Mei}}]{Fan:2020zhy}%
  \BibitemOpen
  \bibfield  {author} {\bibinfo {author} {\bibfnamefont {H.-M.}\ \bibnamefont
  {Fan}}, \bibinfo {author} {\bibfnamefont {Y.-M.}\ \bibnamefont {Hu}},
  \bibinfo {author} {\bibfnamefont {E.}~\bibnamefont {Barausse}}, \bibinfo
  {author} {\bibfnamefont {A.}~\bibnamefont {Sesana}}, \bibinfo {author}
  {\bibfnamefont {J.-d.}\ \bibnamefont {Zhang}}, \bibinfo {author}
  {\bibfnamefont {X.}~\bibnamefont {Zhang}}, \bibinfo {author} {\bibfnamefont
  {T.-G.}\ \bibnamefont {Zi}}, \ and\ \bibinfo {author} {\bibfnamefont
  {J.}~\bibnamefont {Mei}},\ }\href {\doibase 10.1103/PhysRevD.102.063016}
  {\bibfield  {journal} {\bibinfo  {journal} {Phys. Rev. D}\ }\textbf {\bibinfo
  {volume} {102}},\ \bibinfo {pages} {063016} (\bibinfo {year} {2020})},\
  \Eprint {http://arxiv.org/abs/2005.08212} {arXiv:2005.08212 [astro-ph.HE]}
  \BibitemShut {NoStop}%
\bibitem [{\citenamefont {Zi}\ \emph {et~al.}(2023)\citenamefont {Zi},
  \citenamefont {Zhou}, \citenamefont {Wang}, \citenamefont {Li}, \citenamefont
  {Zhang},\ and\ \citenamefont {Chen}}]{Zi:2022hcc}%
  \BibitemOpen
  \bibfield  {author} {\bibinfo {author} {\bibfnamefont {T.}~\bibnamefont
  {Zi}}, \bibinfo {author} {\bibfnamefont {Z.}~\bibnamefont {Zhou}}, \bibinfo
  {author} {\bibfnamefont {H.-T.}\ \bibnamefont {Wang}}, \bibinfo {author}
  {\bibfnamefont {P.-C.}\ \bibnamefont {Li}}, \bibinfo {author} {\bibfnamefont
  {J.-d.}\ \bibnamefont {Zhang}}, \ and\ \bibinfo {author} {\bibfnamefont
  {B.}~\bibnamefont {Chen}},\ }\href {\doibase 10.1103/PhysRevD.107.023005}
  {\bibfield  {journal} {\bibinfo  {journal} {Phys. Rev. D}\ }\textbf {\bibinfo
  {volume} {107}},\ \bibinfo {pages} {023005} (\bibinfo {year} {2023})},\
  \Eprint {http://arxiv.org/abs/2205.00425} {arXiv:2205.00425 [gr-qc]}
  \BibitemShut {NoStop}%
\bibitem [{\citenamefont {Barsanti}\ \emph {et~al.}(2022)\citenamefont
  {Barsanti}, \citenamefont {Franchini}, \citenamefont {Gualtieri},
  \citenamefont {Maselli},\ and\ \citenamefont {Sotiriou}}]{Barsanti:2022ana}%
  \BibitemOpen
  \bibfield  {author} {\bibinfo {author} {\bibfnamefont {S.}~\bibnamefont
  {Barsanti}}, \bibinfo {author} {\bibfnamefont {N.}~\bibnamefont {Franchini}},
  \bibinfo {author} {\bibfnamefont {L.}~\bibnamefont {Gualtieri}}, \bibinfo
  {author} {\bibfnamefont {A.}~\bibnamefont {Maselli}}, \ and\ \bibinfo
  {author} {\bibfnamefont {T.~P.}\ \bibnamefont {Sotiriou}},\ }\href {\doibase
  10.1103/PhysRevD.106.044029} {\bibfield  {journal} {\bibinfo  {journal}
  {Phys. Rev. D}\ }\textbf {\bibinfo {volume} {106}},\ \bibinfo {pages}
  {044029} (\bibinfo {year} {2022})},\ \Eprint
  {http://arxiv.org/abs/2203.05003} {arXiv:2203.05003 [gr-qc]} \BibitemShut
  {NoStop}%
\bibitem [{\citenamefont {Gupta}\ \emph {et~al.}(2021)\citenamefont {Gupta},
  \citenamefont {Bonga}, \citenamefont {Chua},\ and\ \citenamefont
  {Tanaka}}]{Gupta:2021cno}%
  \BibitemOpen
  \bibfield  {author} {\bibinfo {author} {\bibfnamefont {P.}~\bibnamefont
  {Gupta}}, \bibinfo {author} {\bibfnamefont {B.}~\bibnamefont {Bonga}},
  \bibinfo {author} {\bibfnamefont {A.~J.~K.}\ \bibnamefont {Chua}}, \ and\
  \bibinfo {author} {\bibfnamefont {T.}~\bibnamefont {Tanaka}},\ }\href
  {\doibase 10.1103/PhysRevD.104.044056} {\bibfield  {journal} {\bibinfo
  {journal} {Phys. Rev. D}\ }\textbf {\bibinfo {volume} {104}},\ \bibinfo
  {pages} {044056} (\bibinfo {year} {2021})},\ \Eprint
  {http://arxiv.org/abs/2104.03422} {arXiv:2104.03422 [gr-qc]} \BibitemShut
  {NoStop}%
\bibitem{Babak:2009ua}
S.~Babak, J.~R.~Gair and E.~K.~Porter,
Class. Quant. Grav. \textbf{26} (2009), 135004
doi:10.1088/0264-9381/26/13/135004
[arXiv:0902.4133 [gr-qc]].
\bibitem{Babak:2009ua}
S.~Babak, J.~R.~Gair and E.~K.~Porter,
Class. Quant. Grav. \textbf{26} (2009), 135004
doi:10.1088/0264-9381/26/13/135004
[arXiv:0902.4133 [gr-qc]].
\bibitem{Wang:2012xh}
Y.~Wang, Y.~Shang, S.~Babak, Y.~Shang and S.~Babak,
Phys. Rev. D \textbf{86} (2012), 104050
doi:10.1103/PhysRevD.86.104050
[arXiv:1207.4956 [gr-qc]].


\bibitem{Ye:2024bku}
C.~Q.~Ye, H.~M.~Fan, A.~Torres-Orjuela, J.~d.~Zhang and Y.~M.~Hu,
Phys. Rev. D \textbf{109} (2024) no.12, 124034
doi:10.1103/PhysRevD.109.124034

  
  
  
  
  
  
\bibitem{Katz:2021yft}
M.~L.~Katz, A.~J.~K.~Chua, L.~Speri, N.~Warburton and S.~A.~Hughes,
Phys. Rev. D \textbf{104} (2021) no.6, 064047
doi:10.1103/PhysRevD.104.064047
[arXiv:2104.04582 [gr-qc]].
  
\bibitem{Zhang:2022rfr}
C.~Zhang, Y.~Gong, D.~Liang and B.~Wang,
JCAP \textbf{06} (2023), 054
doi:10.1088/1475-7516/2023/06/054
[arXiv:2210.11121 [gr-qc]]. 

\bibitem{Zhang:2023vok}
C.~Zhang, H.~Guo, Y.~Gong and B.~Wang,
JCAP \textbf{06} (2023), 020
doi:10.1088/1475-7516/2023/06/020
[arXiv:2301.05915 [gr-qc]].
\bibitem [{\citenamefont {Datta}\ and\ \citenamefont
  {Phukon}(2021)}]{Datta:2021row}%
  \BibitemOpen
  \bibfield  {author} {\bibinfo {author} {\bibfnamefont {S.}~\bibnamefont
  {Datta}}\ and\ \bibinfo {author} {\bibfnamefont {K.~S.}\ \bibnamefont
  {Phukon}},\ }\href {\doibase 10.1103/PhysRevD.104.124062} {\bibfield
  {journal} {\bibinfo  {journal} {Phys. Rev. D}\ }\textbf {\bibinfo {volume}
  {104}},\ \bibinfo {pages} {124062} (\bibinfo {year} {2021})},\ \Eprint
  {http://arxiv.org/abs/2105.11140} {arXiv:2105.11140 [gr-qc]} \BibitemShut
  {NoStop}%
\bibitem [{\citenamefont {Poisson}(2004)}]{Poisson:2003nc}%
  \BibitemOpen
  \bibfield  {author} {\bibinfo {author} {\bibfnamefont {E.}~\bibnamefont
  {Poisson}},\ }\href {\doibase 10.12942/lrr-2004-6} {\bibfield  {journal}
  {\bibinfo  {journal} {Living Rev. Rel.}\ }\textbf {\bibinfo {volume} {7}},\
  \bibinfo {pages} {6} (\bibinfo {year} {2004})},\ \Eprint
  {http://arxiv.org/abs/gr-qc/0306052} {arXiv:gr-qc/0306052} \BibitemShut
  {NoStop}%
  

  
\end{thebibliography}

%

\end{document}